\documentclass[sigconf,authorversion]{acmart}

\usepackage{booktabs} 

\copyrightyear{\vspace*{40pt}}
\acmBooktitle{Proceedings of the 2019 World Wide Web Conference (WWW '19), May 13--17, 2019, San Francisco, CA, USA}
\acmDOI{10.1145/3308558.3313628}

\usepackage{balance}

\usepackage{amsmath,amssymb,amsfonts,bbm}
\usepackage[ruled,linesnumbered]{algorithm2e}
\usepackage{graphicx}
\usepackage{subfigure}
\usepackage{textcomp}
\usepackage{xcolor}
\usepackage{xspace}
\usepackage{lipsum}
\usepackage{adjustbox}
\usepackage{mathtools}
\usepackage{cuted}

\newcommand{\eg}{\textit{e.g.},\xspace}
\newcommand{\ie}{\textit{i.e.},\xspace}

\DeclareMathOperator*{\argmin}{arg\,min}

\newenvironment{tightlist}{
\begin{list}{$\bullet$}{
    \setlength{\topsep}{2pt}
    \setlength{\partopsep}{0in}
    \setlength{\parskip}{0in}
    \setlength{\itemsep}{0in}
    \setlength{\parsep}{1pt}
    \setlength{\leftmargin}{0in}
    \setlength{\rightmargin}{0in}
    \setlength{\itemindent}{0.12in}
}}
{\end{list}}

\newtheorem{definition}{Definition}[section]
\newtheorem{proposition}{Proposition}[section]
\renewcommand{\comment}[1]{}

\newcommand{\rhov}{\boldsymbol{\rho}}

\newcommand{\Am}{{\bf A}}

\newcommand{\Cm}{{\bf C}}
\newcommand{\Dm}{{\bf D}}

\newcommand{\Mm}{{\bf M}}

\newcommand{\pf}{\noindent{\bf Proof:~}}
\newcommand{\qedsymb}{\hfill{\rule{2mm}{2mm}}}

\begin{document}

\title{Urban Vibes and Rural Charms: Analysis of Geographic Diversity in Mobile Service Usage at National Scale}

\author{Rajkarn Singh}
\orcid{0000-0003-0035-2471}
\affiliation{%
  \institution{The University of Edinburgh}
  \streetaddress{10 Crichton Street}
  \city{Edinburgh}
  \state{UK}
  \postcode{EH8 9AB}
}
\email{r.singh@ed.ac.uk}

\author{Marco Fiore}
\orcid{0000-0002-0772-9967}
\affiliation{%
  \institution{CNR-IEIIT}
  \streetaddress{Corso Duca degli Abruzzi 24}
  \city{Turin}
  \state{Italy}
  \postcode{10129}
}
\email{marco.fiore@ieiit.cnr.it}

\author{Mahesh K. Marina}
\affiliation{%
  \institution{The University of Edinburgh}
  \streetaddress{10 Crichton Street}
  \city{Edinburgh}
  \state{UK}
  \postcode{EH8 9AB}
}
\email{mahesh@ed.ac.uk}

\author{Alessandro Nordio}
\orcid{0000-0001-8258-051X}
\affiliation{%
  \institution{CNR-IEIIT}
  \streetaddress{Corso Duca degli Abruzzi 24}
  \city{Turin}
  \state{Italy}
  \postcode{10129}
}
\email{alessandro.nordio@ieiit.cnr.it}

\author{Alberto Tarable}
\affiliation{%
  \institution{CNR-IEIIT}
  \streetaddress{Corso Duca degli Abruzzi 24}
  \city{Turin}
  \state{Italy}
  \postcode{10129}
}
\email{alberto.tarable@ieiit.cnr.it}

\renewcommand{\shortauthors}{R. Singh et al.}

\begin{abstract}
We investigate spatial patterns in mobile service consumption that emerge at national scale. Our investigation focuses on a representative case study, \ie France, where we find that: ($i$) the demand for popular mobile services is fairly uniform across the whole country, and only a reduced set of peculiar services (mainly operating system updates and long-lived video streaming) yields geographic diversity; ($ii$) even for such distinguishing services, the spatial heterogeneity of demands is limited, and a small set of consumption behaviors is sufficient to characterize most of the mobile service usage across the country; ($iii$) the spatial distribution of these behaviors correlates well with the urbanization level, ultimately suggesting that the adoption of geographically-diverse mobile applications is linked to a dichotomy of cities and rural areas. We derive our results through the analysis of substantial measurement data collected by a major mobile network operator, leveraging an approach rooted in information theory that can be readily applied to other scenarios.
\end{abstract}

%
%



\maketitle

\section{Introduction}
\label{sec:intro}

As mobile data traffic keeps surging worldwide~\cite{cisco17}, knowledge of where, when, how and why mobile services are consumed by network subscribers becomes increasingly relevant across research and technology domains,
including sociology~\cite{blondel08}, demography~\cite{denadai16}, urban planning~\cite{grauwin15}, economy~\cite{smith13}, transportation engineering~\cite{zhang14}, or network management~\cite{marquez18}.
Despite the importance of the problem and some recent efforts discussed in Section\,\ref{sec:related}, our comprehension of mobile service adoption is currently limited, and many questions remain unanswered, especially when considering the phenomenon at very large geographical scales.
In this paper, we focus on one such open question, namely: \emph{``how similar (or different) are demands for mobile services across a whole country?''} We answer by analyzing a real-world dataset of mobile network traffic collected by a major operator that describes the demands for individual services in 10,000 \emph{communes} (\ie administrative areas) in France. Our study yields the following insights:
\begin{tightlist}
	\item what sets communes apart are not usage patterns of the most popular services, which tend to be similar everywhere in the country, but those of a small set of specific services that still figure in the top-50 services in terms of generated traffic, including operating system updates and long-lived video streaming;
	\item just $9$ (respectively, $50$) service consumption patterns are sufficient to retain $23\%$ (respectively, $35$\%) of the overall usage diversity, implying that a small number of distinct behaviors is sufficient to characterize the many thousands of areas in the whole of France;
	\item clear correlations exist between different patterns in mobile service consumption and demographics features, which are rooted in higher (or lower) than average usage of specific types of service.
\end{tightlist}

Deriving these results requires overcoming methodological and computational challenges. We face a clustering problem, where communes are to be grouped based on how their inhabitants use mobile services. However, our clustering operates in a multidimensional space of hundreds of mobile services, where a suitable notion of similarity is to be defined. In addition, working at a national scale implies potentially disentangling billions of pairwise relationships between tens of thousands of geographical areas.

We address these issues by adopting an information theoretic approach that builds on the notion of \emph{mutual information} between mobile service demands and geographical locations. The mutual information measures how much can be inferred
about the consumed services by knowing the location, and vice versa. We first leverage it to limit the problem dimension in the mobile service space, by identifying \emph{informative services} that maximize the mutual information, \ie exhibit significant spatial diversity. Also, we measure the similarity of usage distributions of such informative services between two communes in terms of the loss of mutual information incurred when their distributions are merged. Finally, the fraction of retained mutual information is used to assess the quality of clustering results obtained via a scalable two-phase approach.

Overall, our work sheds light on the limited set of services that are responsible for diversity in mobile data demands across a whole developed country, and on their relation to demographics features. It also provides the research community with a tool%
\footnote{Available at \url{https://github.com/rajkarn/mobdiv}.}
for the analysis of patterns in mobile data traffic usage at national scales. 

\section{Related Work}
\label{sec:related}

The vast majority of the literature on mobile network traffic analysis investigates patterns in the aggregate demand, without differentiating among services. In this context, early works have revealed the heterogeneity that characterizes the offered load at radio access in space and time, leading to strong fluctuations of the demand across diverse regions of a same city and during different periods of the day~\cite{paul11}. Especially relevant to our study are works that established links between the temporal dynamics of aggregate traffic and the land use, \ie the type of human infrastructures and activities present in a given area~\cite{cici15,furno17,xu17}. Although they focus on aggregate traffic at city scale, these studies have demonstrated for the first time the strong impact that user centric aspects can have on the adoption of mobile applications.

At the individual mobile service level, several works have studied specific applications in depth. Investigations have focused on services such as adaptive-bit-rate video streaming by over-the-top providers~\cite{erman11}, Facebook and WhatsApp~\cite{fiadino15}, or cloud storage~\cite{li16}. These studies address the network-level performance of the examined services, and do not provide insights in terms of the geographical diversity of their usage. Related to these works is also a thorough analysis of web browsing behaviors by mobile users, which finds that a limited number of profiles are sufficient to capture most of the patterns in website visits~\cite{keralapura10}. However, similarly to the papers above, also in this case the spatial dimension is not considered.

Geographical diversity has been often overlooked also in prior explorations of mobile data traffic considering multiple services. Most works in the literature have a different focus, including differences in mobile application usage in time~\cite{zhang12} or across the subscriber population~\cite{li15}. Attention has also been paid to patterns in the utilization of apps by individual users, finding, \eg that mobile service usage is very heterogeneous among the user base~\cite{falaki10}, strongly depends on context~\cite{bohmer11}, is characterized by brief bursts of interactions~\cite{ferreira14}, and is influenced by the type of device used~\cite{hintze17}. However, these are orthogonal problems to that of the spatial diversity of the demands for mobile services that we target. Finally, several previous works have observed a strong locality in the usage of applications within a given urban area, \ie a significant spatial diversity of usage across different city neighborhoods~\cite{trestian09,shafiq12}. Our investigation suggests that these dissimilarities in service usage are not significant at a national scale, where the consumption of mobile services is relatively uniform.

Only two previous works investigate the spatial dimension of mobile application usage at a national scale. In a study carried out in the US~\cite{xu11}, the authors hint at the existence of local (\ie US state-related) and nationwide services. We do not find such a dichotomy in our case study, and ascribe it to the federal organization of the US into states, which is not reflected in France. Interestingly, however, the differences in the consumption of nationwide apps is fairly limited in the USA (between 2\% and 20\%), which is consistent with our findings.
The second work is a recent study of mobile service usage in France~\cite{marquez17}, which unveils the existence of a strong temporal diversity in mobile service demands, but somewhat lower geographical differences. Our in-depth analysis substantiates the observations in~\cite{marquez17} with stronger evidence of the limited nationwide diversity of mobile service consumption patterns.  
Finally, we remark that both works above only provide preliminary geographical results via baseline statistical measures, and do not perform a rigorous analysis based on spatial clustering as the one we propose. Moreover, none investigates the existence of informative services whose distribution is geographically varied, or offers interpretations of the results based on side information.

\section{System model}
\label{sec:model}

As stated at the outset, our framework builds on information theory. In this section we introduce the notation and fundamentals of the proposed approach, and show how they apply to our scenario.

\subsection{Probabilistic rendering of service demands}
\label{sub:prob}

Let ${\mathcal C}=\{ 1,\ldots,N_C\}$ and ${\mathcal S}=\{ 1,\ldots,N_S\}$ be the set of geographical areas in the target region and the set of mobile services under study, respectively, having cardinality $N_C$ and $N_S$. A total mobile data traffic of $t_i$ bytes is generated in area $i$, $i=1,\ldots,N_C$, over the system observation time, \ie the time period during which network measurements are performed%
\footnote{As our interest is in the spatial diversity of mobile service usage, we primarily consider data accumulated over time. However, we also carried out experiments by partially disaggregating data in time, as discussed at the end of Section\,\ref{subsec:linking}.}.
Let $C$ be a random variable, with outcome in ${\mathcal C}$, representing the selection of an area with a certain probability. Similarly, let $S$ be a random variable, with outcome in ${\mathcal S}$, representing the selection of a service.
In the considered period of time, the probability that a given byte of traffic was generated by service $j$ in area $i$ is the joint probability distribution of services and areas, denoted by $p_{S,C}(j,i)$, $j=1,\ldots,N_S$, $i=1,\ldots,N_C$.

Given an area $i$, the probability of observing a byte generated by service $j$ is denoted by the conditional probability $p_{S|C}(j|i)=\rho_{j,i}$. The vector $\boldsymbol{\rho}_i=[\rho_{1,i},\ldots,\rho_{N_S,i}]$ thus represents the {\em service usage distribution} in area $i$ and is such that $\sum_{j=1}^{N_S}\rho_{j,i}=1$.

Overall, the probability of observing traffic generated by service $j$, $j=1,\ldots,N_S$, is represented by the marginal distribution of traffic among services:
\[ p_S(j) = \sum_{i=1}^{N_C} p_{S,C}(j,i) = \sum_{i=1}^{N_C} p_C(i) \rho_{j,i}, \]
where $p_C(i)$, $i=1,\ldots,N_C$ is the fraction of traffic in area $i$, among all areas, and so is given by: 
\[ p_C(i) = \frac{t_i}{\sum_{k=1}^{N_C}t_k}.\]

The amount of information that random variables $S$ and $C$ share is measured by the \textit{mutual information}:
\begin{equation}
\label{eqnISC}
I(S;C) = H(S)-H(S|C),
\end{equation}
where $H(S)$ and $H(S|C)$ are the entropy of $S$ and the conditional entropy of $S$ given $C$, respectively, expressed as:
\[ H(S) = - \sum_{j=1}^{N_S} p_S(j) \log p_S(j), \]
and
\[ H(S|C) = -\sum_{i=1}^{N_C} p_C(i) \sum_{j=1}^{N_S} p_{S|C}(j|i) \log p_{S|C}(j|i). \]

The mutual information $I(S;C)$ is a measure of the correlation between how traffic is distributed among services in $S$ and among areas in $C$; \ie it captures how much can be inferred about the consumed services by knowing $C$, or vice versa. As a result, $I(S;C)$ measures how much mobile service usage depends on geographical location. $I(S;C)=0$ when $S$ and $C$ are independent, \ie the exact same distribution of mobile service traffic is observed across all areas in $C$, and the spatial diversity is nil. Non-zero yet low values of mutual information of services and areas imply that sampling a unit of traffic from those services still carries little information on the area it was sampled from, \ie that services tend to have a quite uniform usage distribution across areas. As $I(S;C)$ grows, the knowledge of $C$ increasingly helps to anticipate the value of $S$, \ie geographical regions are characterized by more and more distinctive mobile service usage, hence the spatial diversity rises.

\subsection{Application to the France case study}

\begin{figure*}[tb]
	\centering
	\includegraphics[height=0.35\columnwidth]{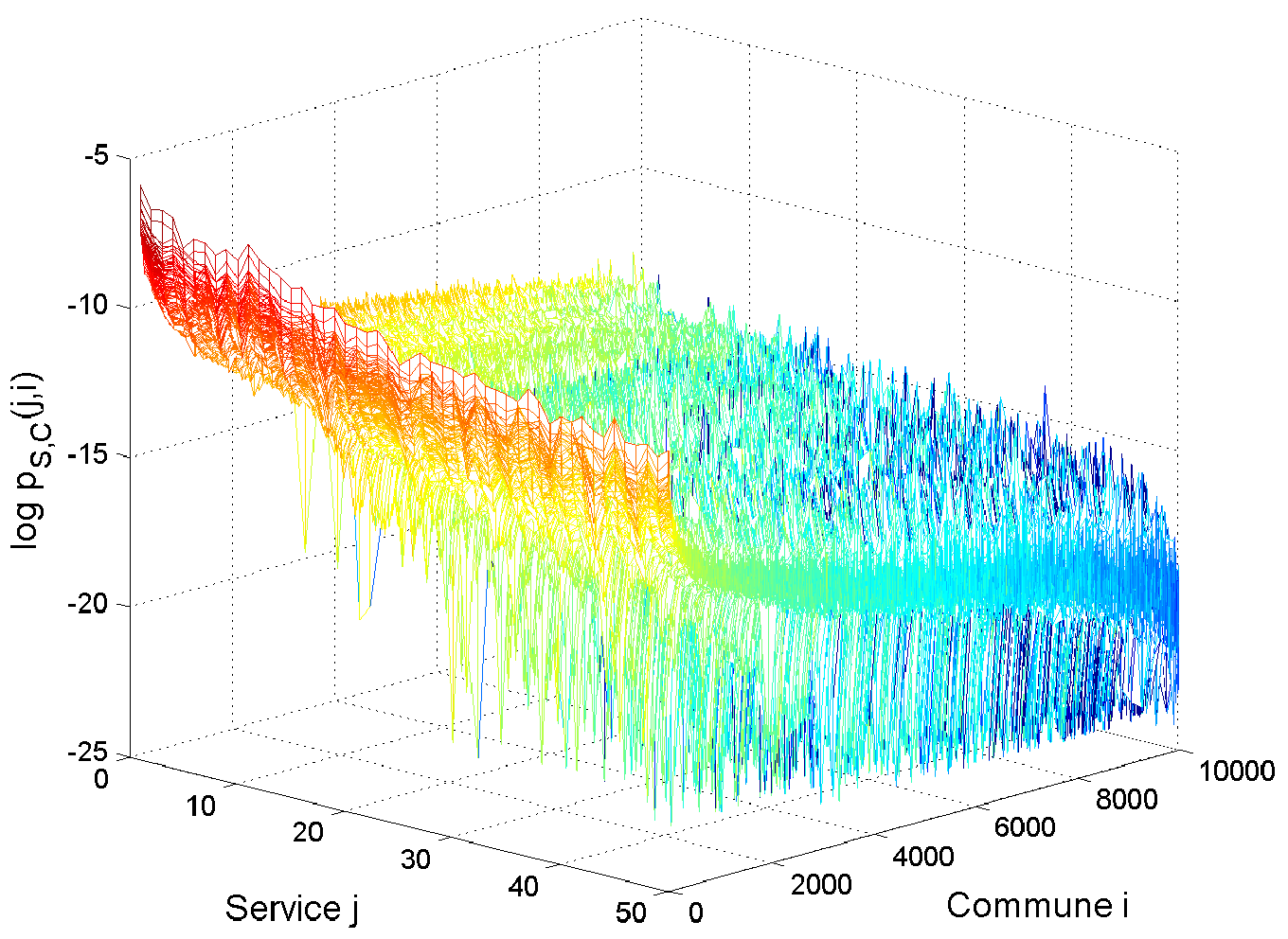}
	\hspace*{5pt}
	\includegraphics[height=0.33\columnwidth]{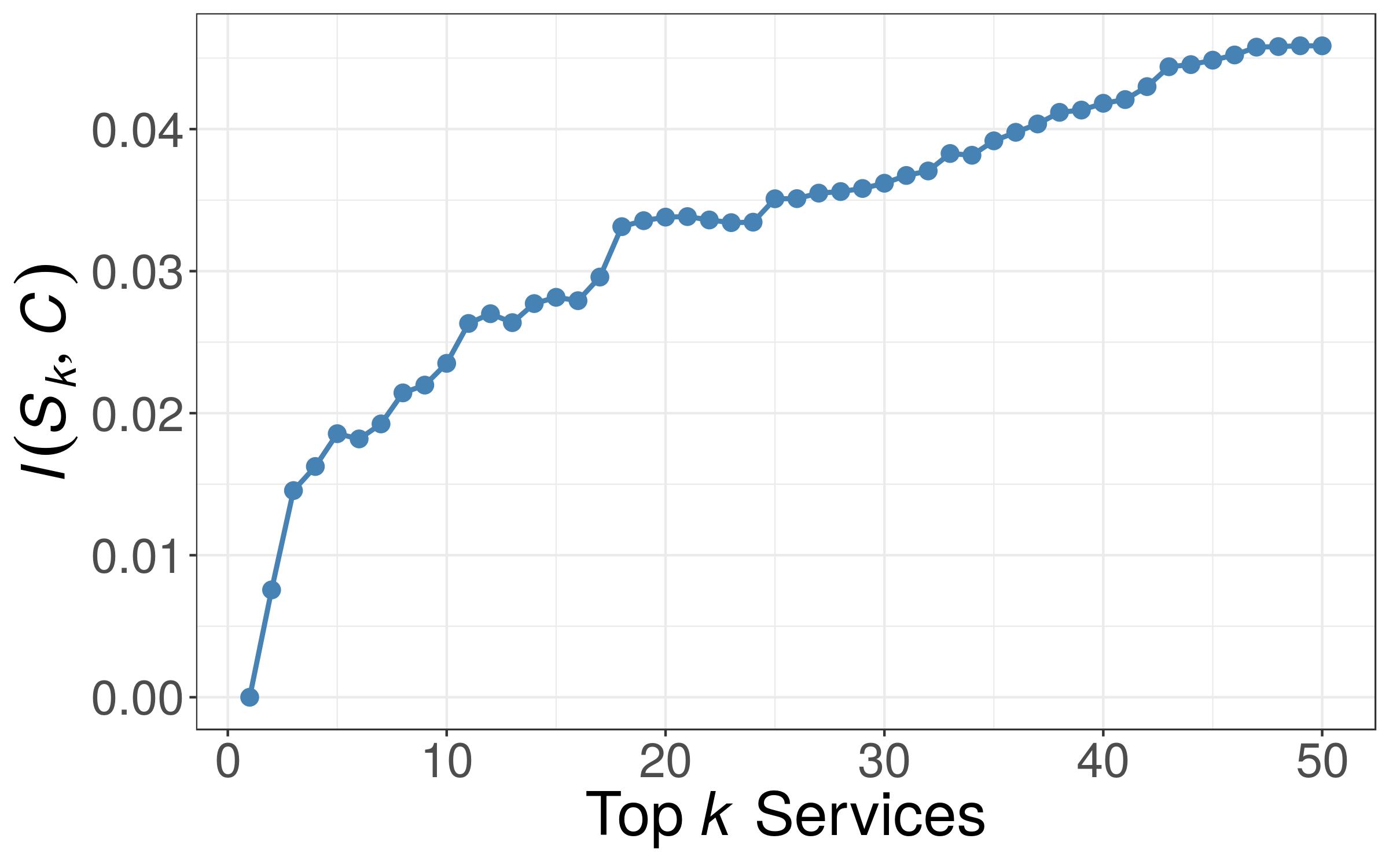}
	\hspace*{10pt}
	\includegraphics[height=0.34\columnwidth]{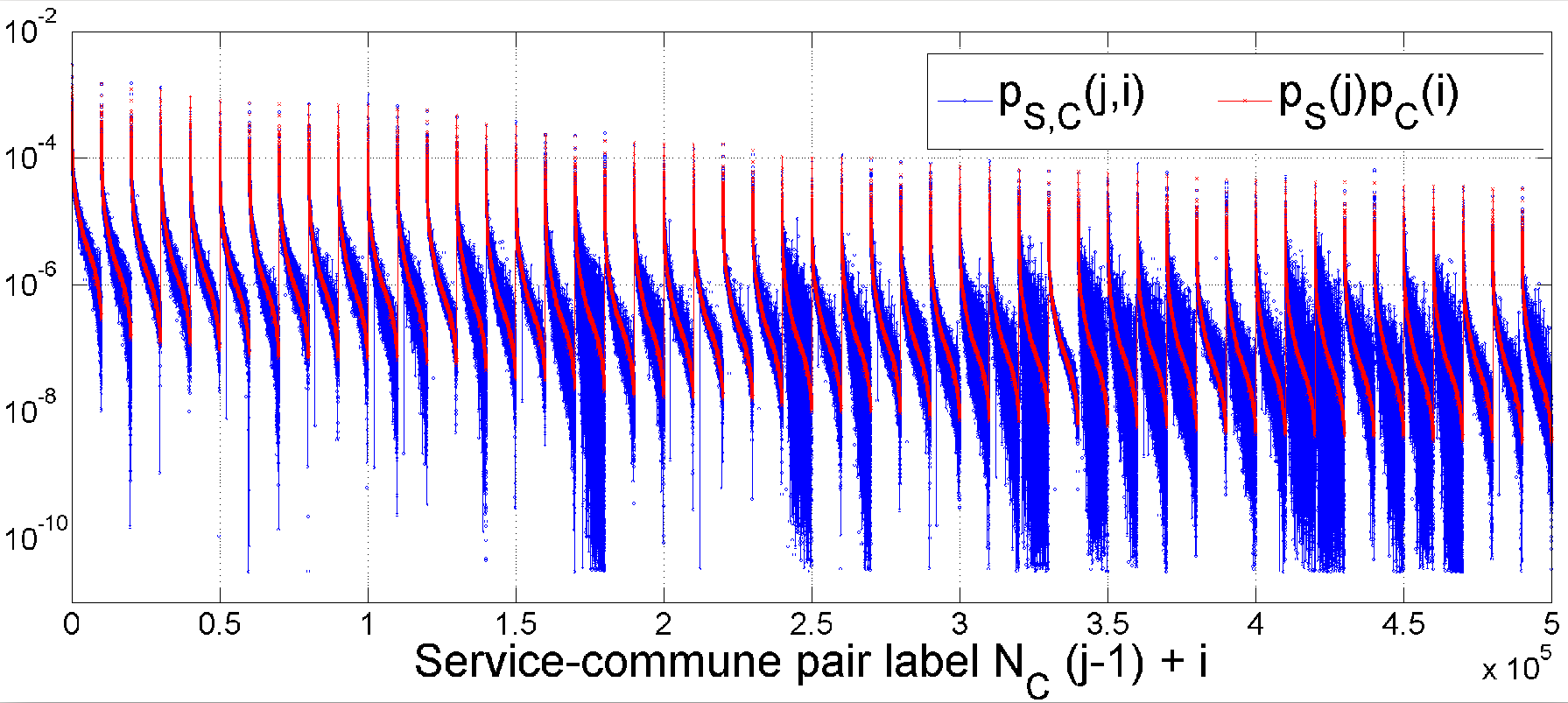}
	\vspace*{-4pt}
	\caption{Probabilistic view of mobile service demands in France. Left: joint distribution $p_{S,C}(j,i)$ of services in $S$ and communes in $C$. Middle: mutual information $I(S_k;C)$ computed on subsets $\mathcal{S}_k$ of the $k$ services generating the most traffic. Right: unidimensional rendering of $p_{S,C}(j,i)$ (blue) and marginal distributions product $p_S(j) \cdot p_C(i)$ (red). Figure best viewed in colors.}
	\label{fig:prob}
	\vspace*{-8pt}
\end{figure*}

We apply the probabilistic model above to the nationwide case study of France. The information on the traffic demands for individual mobile services in France was collected by a major network operator during one continuous week in late 2016. Deep packet inspection and proprietary fingerprinting techniques%
\footnote{Confidential agreements with the network operator do not allow us to disclose the details of the traffic classification procedure. However, we can mention that 88\% of the sessions were correctly detected, according to the operator's performance evaluations. Also, we underscore that the data collection occurred in compliance with regulations in force, and was approved by the French national authority for data privacy (CNIL). All data was aggregated by the operator at the commune level before we could access it, which ensures strong privacy protection as mobile traffic is accumulated over thousands of subscribers. No individual data is used in our study.}
were employed on traffic sniffed on the GPRS Tunneling Protocol (GTP), so as to associate single IP-level flows to applications.
Overall, the dataset captures the usage of thousands of mobile services by 30 million subscribers in 10,000 \textit{communes}, \ie local administrative zones with a mean surface of 16 km\textsuperscript{2}, in France.

The left plot in Fig.\,\ref{fig:prob} provides an illustration of the joint probability $p_{S,C}(j,i)$, where $\mathcal{S}$ is the set of the 50 most popular mobile services%
\footnote{We limit plots to the 50 services that generate the highest demands, for the sake of clarity. Such services account for over 93\% of the total mobile data traffic in France.},
and $\mathcal{C}$ is the set of 10,000 communes. We remark the high unbalance in the traffic recorded across communes, as well as among services, underscored by the logarithmic scale of the z axis.

The middle plot in Fig.\,\ref{fig:prob} shows the evolution of the mutual information $I(S_k;C)$, computed by limiting the joint distribution to a subset $\mathcal{S}_k \subseteq \mathcal{S}$ of the $k$ services that generate the highest total traffic; equivalently, $S_k$ denotes the random variable representing the selection of a service in $\mathcal{S}_k$. The number $k$ of considered services is on the x-axis. We note that the mutual information remains very low, close to zero, for all top-50 services: according to our previous discussion, this implies that there is very little correlation between the two variables, hence the distribution of mobile service traffic does not vary in a sensible manner across communes.

A more detailed view is provided in the right plot in Fig.\,\ref{fig:prob}, which shows the joint distribution $p_{S,C}(j,i)$ in a linearized form where each ``period'' represents the probabilities associated to one service over all communes. The product of the marginal distributions $p_S(j) \cdot p_C(i)$ is also displayed according to the same format. There is a substantial overlap between the two curves, which confirms that spatial diversity is low also when inspecting the system on a more detailed per-service basis. Indeed, matching curves imply $p_{S,C}(j,i) = p_S(j) \cdot p_C(i)$, hence independence between $S$ and $C$, or, equivalently, $I(S;C)=0$ in \eqref{eqnISC}. As discussed previously, such a condition indicates that service consumption is identical everywhere.

Overall, these results lead to our first takeaway message: \textit{\textbf{usage patterns of the most popular mobile services tend to be very similar across the whole country under study}}. This is quite a surprising outcome, considering the spatial differences in demographic, social and economic features that characterize France. It also leads us to investigate next if specific individual services are in fact geographically diverse in their usage characteristics.

\section{Detecting informative services}
\label{sec:services}

An interesting observation from the right plot of Fig.\,\ref{fig:prob} is that, although the overlap between $p_{S,C}(j,i)$ and $p_S(j) \cdot p_C(i)$ is generally good, some services (denoted by specific ``periods'') show especially noisy joint distribution curves that are not well captured by the simple product of the marginal distributions. Such services thus appear to be adopted less homogeneously across the country than most other mobile applications.
Next, we investigate the existence of \emph{informative services} that are characterized by a non-negligible diversity of usage across geographical areas in the target region.

\subsection{Maximizing the mutual information}

Consider a subset $\mathcal{S}' = \{j_1,\dots,j_{|\mathcal{S}'|} \} \subseteq \mathcal{S}$ of services and define the service usage distribution in area $i$ restricted to $\mathcal{S}' $ as $\boldsymbol{\rho}_i (\mathcal{S}') = [\rho_{i1}(\mathcal{S}'), \dots, \rho_{i,|\mathcal{S}'|}(\mathcal{S}')]$, where
\[
\rho_{ik}(\mathcal{S}') = \frac{\rho_{i,j_k}}{\sum_{k'=1}^{|\mathcal{S}'|} \rho_{i,j_{k'}}}.
\]

Considering only services that are in $\mathcal{S}'$, let us define the traffic within area $i$, denoted by $t_{i}(\mathcal{S}')$, as the sum of the demands within area $i$ for each service in $\mathcal{S}'$.  The joint probability of sampling area $i$ and service $j_k \in \mathcal{S}'$ is then given by $p_{C}(i, \mathcal{S}') \rho_{ik}(\mathcal{S}')$, where we suppose $p_{C}(i, \mathcal{S}')$ to be the fraction of traffic in area $i$, \ie 
\[
p_{C}(i, \mathcal{S}') = \frac{t_{i}(\mathcal{S}') }{ \sum_{i'=1}^{N_C}  t_{i'}(\mathcal{S}')}.
\]

We can now define informative services as a subset of all mobile services such that their mutual information with respect to the target spatial areas is maximized. Formally, the subset of informative services defined as above maps to the optimal choice of $\mathcal{S}'\subseteq\mathcal{S}$ that solves the following problem:
\[
\mathcal{S}'_{\mathrm{opt}} = \arg \max_{\mathcal{S}' \subseteq \mathcal{S}} I(C |_{\mathcal{S}'}; S|_{\mathcal{S}'}),
\] 
where $C |_{\mathcal{S}'}$ and $S |_{\mathcal{S}'}$ are the random variables representing the sampled area and service in the restricted scenario where only services in $\mathcal{S}'$ are considered.

Unfortunately, the above combinatorial problem is  too complex to be solved exactly for typical data sizes. We thus adopt the following heuristic approach to determine $\mathcal{S}'$. Let us consider the whole set of services and sample service $j$ from an arbitrary probability distribution $p_S(j)$. Then, along the lines of \eqref{eqnISC}, we consider $I(C;S) = H(S)-H(C|S)$, where 
\[
H(C|S) = -\sum_{j=1}^{N_S} p_S(j) \sum_{i=1}^{N_C} p_{C|S}(i|j) \log p_{C|S}(i|j),
\]
and area $i$ is sampled with probability
\[ \label{eq:Bayes}
p_{C|S}(i|j) = \frac{t_{i,j}}{\sum_{i'=1}^{N_C} t_{i',j}}.
\]
The optimal service distribution $p_S^*$ that weights services according to their informativeness is the one that solves
\[
p_S^* = \arg \max_{p_S(j)} I(C;S),
\]
which we obtain via the Blahut-Arimoto algorithm~\cite{Arimoto,Blahut}.
We then sort the services according to decreasing values of $p_S^*$ and we set a threshold $\theta$. Then, $\mathcal{S}'$ contains those services for which $p_S^* > \theta$. The value of $\theta$ can be empirically chosen by inspecting the shape of $p_S^*$, as we will see next in the context of our reference scenario.

\subsection{Application to the France case study}
\label{sub:informative-france}

\begin{figure}[tb]
	\centering
	\includegraphics[width=0.5\columnwidth]{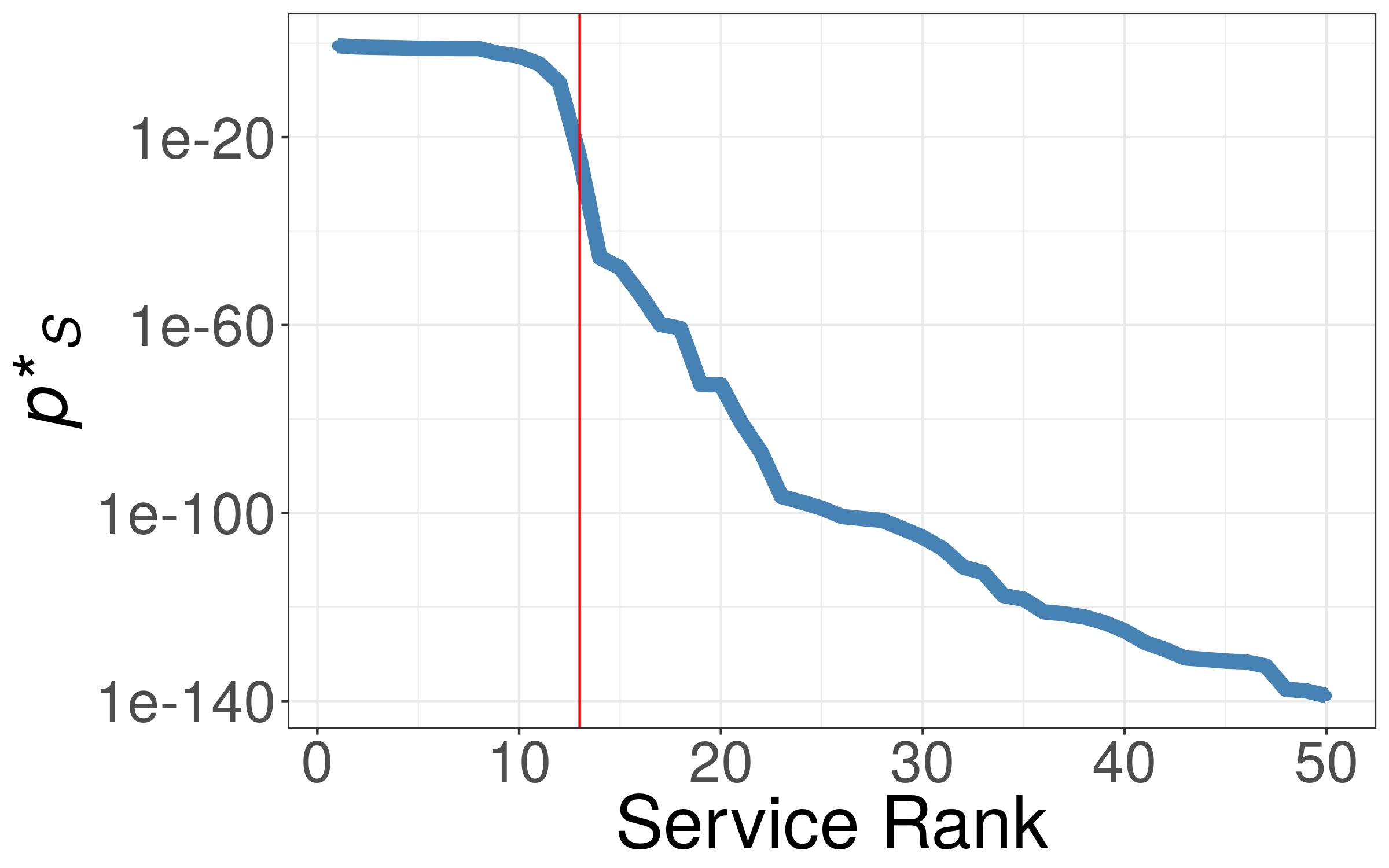}
	\vspace*{-6pt}
	\caption{Mobile services ranked by the weighting distribution $p_S^*$ returned by the Blahut-Arimoto approach in France.}
	\label{fig:arimoto}
	\vspace*{-8pt}
\end{figure}

When applied to the France reference scenario, the approach returns the result in Fig.~\ref{fig:arimoto}. Here, mobile services in $\mathcal{S}$ (x axis) are ranked based on their associated weights $p_S^*$. The differences in weights is striking, and spans tens of orders of magnitude: this implies that some services are significantly more informative than others.
Specifically, there is a clear gap in the ranked values after the 13\textsuperscript{th} service, highlighted by the vertical line, which indicates a substantial reduction of informativeness beyond that point.
As the first 13 mobile services in the $p_S^*$ ranking yield nearly equivalent informativeness, we include them all in the set $\mathcal{S}'$. These informative services are listed in Tab.\,\ref{tab:services}, where we remark that:
\begin{tightlist}
	\item there exists a notable pattern in the selected mobile services, as they belong to specific classes, \ie operating system (OS) updates, and audio/video streaming;
	\item none of these informative services are among the top 15 in terms of total generated traffic, but they are all within the top 50, and amount to substantial network traffic loads of Terabytes per week.
\end{tightlist}

\begin{table}[tb]
	\centering
	\caption{Set $\mathcal{S}'$ of informative services in France. Values within parentheses denote OS-specific traffic, while the absence of such values indicates OS-independent traffic.}
	\vspace*{-4pt}
	\fontsize{7}{2}\selectfont%
	\renewcommand{\arraystretch}{4}
	\setlength{\tabcolsep}{6pt}
	\begin{tabular}{|r|l|rl|}
		\hline
		{\it $p_S^*$ rank} & {\it Service name} & {\it Weekly traffic} & {\it/ Overall rank}\\
		\hline
		\hline
		1 & torrent (Android) & 8,070 GB & {\it/} 43 \\
		2 & Netflix (Android) & 8,244 GB & {\it/} 42 \\
		3 & Netflix (iOS) & 13,100 GB & {\it/} 33 \\
		4 & streaming (Android) & 6,989 GB & {\it/} 47 \\
		5 & OS updates (iOS) & 10,100 GB & {\it/} 38 \\
		6 & streaming (Windows Mobile) & 11,800 GB & {\it/} 35 \\
		7 & streaming (iOS) & 20,503 GB & {\it/} 25 \\
		8 & OS updates (Windows Mobile) & 42,900 GB & {\it/} 18 \\
		9 & WhatsApp & 8,821 GB & {\it/} 40 \\
		10 & OS updates (Android) & 6,993 GB & {\it/} 46 \\
		11 & blogging & 11,100 GB & {\it/} 37 \\
		12 & cloud storage (iOS) & 11,300 GB & {\it/} 36 \\
		13 & SoundCloud & 7,814 GB & {\it/} 45 \\
		\hline
	\end{tabular}
	\label{tab:services}
	\vspace*{-8pt}
\end{table}

Overall, those identified above are the mobile services that show substantial diversity in usage across France. Indeed, if $S'$ is the random variable representing the selection of a service in $\mathcal{S}'$, the mutual information $I(S';C)$ is equal to 0.236, which is between 5 and 25 times that recorded from the top-$k$ services in $\mathcal{S}_k$, for any $k$$\leq$$50$, in the middle plot of Fig.\,\ref{fig:prob}. The findings above convey our second takeaway message: \textit{\textbf{there exist a small set of applications that are actually informative of the geographical diversity in mobile service consumption, and which belong to fairly specific service categories.}}

\section{Clustering mobile service usages}
\label{sec:clustering}

In order to understand how the informative services in $\mathcal{S}'$ are linked to the actual geography of France, we cluster the communes in $\mathcal{C}$ based on their mobile service usage distribution. Clustering based on distributions itself is non-trivial, and the scale of our scenario adds a layer of complexity. We build upon a recent breakthrough in social segregation analysis~\cite{chodrow17} to define the \emph{weighted diversity} of two distributions, and use it in a scalable two-phase process.

\subsection{Weighted diversity}
\label{sub:diversity}

\begin{figure*}[tb]
\fcolorbox{white}{white}{
	\begin{minipage}[t][1\height][c]{\dimexpr\textwidth-2\fboxsep-2\fboxrule\relax}
		\begin{align} \label{eq:diffMI}
		I(K_1;S) - I(K_2;S) & = H(S | K_2) - H(S | K_1) \nonumber \qquad \qquad \qquad \qquad \\
		& = \sum_{k = 1}^{N_K} p_{K_2}(k)  H(S | K_2 = k) - \sum_{k = 1}^{N_K+1} p_{K_1}(k)  H(S | K_1 = k) \nonumber \qquad \qquad \qquad \qquad \\
		& = p_{K_2}(N_K) H(S | K_2 = N_K) - p_{K_1}(N_K) H(S | K_1 = N_K)  - p_{K_1}(N_K+1) H(S | K_1 = N_K+1) \qquad \qquad \qquad \qquad
		\end{align}
		\begin{equation} \label{eq:diff_ent}
		H(S | K_2 = N_K) \geq \frac{p_{K_1}(N_K)}{p_{K_1}(N_K) + p_{K_1}(N_K+1)}  H(S | K_1 = N_K)
		+ \frac{p_{K_1}(N_K+1)}{p_{K_1}(N_K) + p_{K_1}(N_K+1)} H(S | K_1 = N_K+1) \qquad \qquad \qquad \qquad
		\end{equation}
	\end{minipage}%
	}
\end{figure*}

Our problem essentially involves clustering geographical areas according to their service usage distribution. In our context, a clustering of areas with $N_K$ clusters ($N_K \leq N_C$) is a map $\mathfrak{K}: \mathcal{C} \rightarrow \mathcal{K}$ where $\mathcal{K} = \{ 1,\dots, N_K\} $. We then define $J_{\mathfrak{K}}(k)$, $k = 1,\dots, N_K$ as 
\[
J_{\mathfrak{K}}(k) = \{ c \in \mathcal{C} : \mathfrak{K}(c) = k\},
\]
which is the set of geographical areas grouped into cluster $k$.
Also, a clustering $\mathfrak{K}_1$ is a \emph{refinement} of another clustering $\mathfrak{K}_2$ if and only if, for any two $c_1, c_2 \in \mathcal{C}$, $\mathfrak{K}_1(c_1) = \mathfrak{K}_1(c_2)$ implies $\mathfrak{K}_2(c_1) = \mathfrak{K}_2(c_2)$.   

Let $K = \mathfrak{K}(C)$ be the random variable representing the cluster in which a sampled area belongs to. The distribution $p_K$ of clusters is induced by the distribution of areas $p_C$ in a trivial way, \ie
\[
p_K(k) = \sum_{i \in J_{\mathfrak{K}}(k)} p_C(i).
\] 
We can now define the mutual information between cluster and service random variables, denoted by $I(K,S)$. By the data processing inequality, we have $I(K;S) \leq I(C;S)$, since $K$ is a deterministic function of $C$ so that the knowledge of $C$ implies the knowledge of $K$ but not vice versa. More generally, we have the proposition:

\begin{proposition}
	If $\mathfrak{K}_1$ is a refinement of $\mathfrak{K}_2$ and $K_i = \mathfrak{K}_i(C)$, $i=1,2$,  then 
	\[
	I(K_2;S) \leq I(K_1;S)
	\]
\end{proposition}

\comment{\begin{proposition}
		If $\mathfrak{K}_2$ is a refinement of $\mathfrak{K}_1$ and $K_i = \mathfrak{K}_i(C)$, $i=1,2, \dots, N_C$,  then 
		\[
		I(K_2;S) \leq I(K_1;S)
		\]
	\end{proposition}
}

\pf
For the proof, we first consider that $\mathfrak{K}_1$ has $N_K+1$ clusters, $\mathfrak{K}_2$ has $N_K$ clusters, $J_{\mathfrak{K}_1}(k) = J_{\mathfrak{K}_2}(k)$ for $k = 1,\dots, N_K-1$ and $J_{\mathfrak{K}_2}(N_K) = J_{\mathfrak{K}_1}(N_K) \cup J_{\mathfrak{K}_1}(N_K+1) $. Then, the difference $I(K_1;S) - I(K_2;S) $ is given in \eqref{eq:diffMI}, where $p_{K_2}(N_K) = p_{K_1}(N_K) + p_{K_1}(N_K+1)$. Now, since mutual information is convex, we have as a consequence \eqref{eq:diff_ent}. This implies that $I(K_1;S) - I(K_2;S) \geq 0$.

When $\mathfrak{K}_1$ is a general refinement of $\mathfrak{K}_2$, we can always imagine a chain of refinements that starts from $\mathfrak{K}_2$ and ends to $\mathfrak{K}_1$, in which each step consists in splitting a cluster into two. For each of such steps, we can apply the argument above, and conclude again that $I(K_1;S) - I(K_2;S) \geq 0$. 
\qedsymb
\vspace*{4pt}

In the above proposition, the information loss $I(K_1;S) - I(K_2;S)$ can be seen as the price to pay for merging two clusters of $\mathfrak{K}_1$ into a single cluster of $\mathfrak{K}_2$. Based on this observation, and given a clustering $\mathfrak{K}$, we can introduce a pairwise cost measure for the joining of any two clusters $J_{\mathfrak{K}}(k_1)$ and $J_\mathfrak{K}(k_2)$, which will be central to the design of our information theory-based clustering algorithm. We name such a measure \emph{weighted diversity}, and define it as follows. 
\begin{definition}
	Given a spatial clustering $\mathfrak{K}$ with clusters $J_\mathfrak{K}(1),$ $\dots,$ $J_\mathfrak{K}(N_K)$, let us define a new clustering $\mathfrak{K}_{i,i'}$, with $N_K-1$ clusters, obtained from $\mathfrak{K}$ by merging clusters $J_\mathfrak{K}(i)$ and  $J_\mathfrak{K}(i')$. Moreover, let $K = \mathfrak{K}(C)$ and $K_{i,i'} = \mathfrak{K}_{i,i'}(C)$. We define the weighted diversity between $J_\mathfrak{K}(i)$ and  $J_\mathfrak{K}(i')$ as
	\begin{equation}
	d(i,i')  =  I(K;S) - I(K_{i,i'}; S).
	\label{eq:div}
	\end{equation}
\end{definition}

In the above proposition, the information loss $I(K_1;S) - I(K_2;S)$ can be seen as the cost entailed by passing from a more refined description of service usage distribution given by clustering $\mathfrak{K}_1$ to a coarser description corresponding to $\mathfrak{K}_2$.
Next, we employ the measure in \eqref{eq:div} as the basis for a clustering algorithm.

\subsection{Practical clustering algorithm}
\label{sub:algo}

The weighted diversity measure can be leveraged as a distance metric for practical algorithms that aim at clustering geographical areas based on mobile service usage. It has the following desirable features: ($i$) it depends only on the two considered clusters $i$ and $i'$, and not on the other components of clustering $\mathfrak{K}$, as per \eqref{eq:diffMI}, hence it is a suitable distance metric for clustering; ($ii$) its definition and properties are independent of how $\mathfrak{K}$ is obtained, hence it can be used in combination with any clustering algorithm; and, ($iii$) it is specifically designed for computing the dissimilarity of two distributions, \ie the data representation that characterizes our system. In addition, the metric has a clear interpretation in information theory terms, as it maps to the loss of information incurred by merging the mobile service usage distributions of two geographical areas into one. Two areas with identical service usage will be characterized by a null weighted diversity, and merging them into the same cluster will preserve the original distributions without any loss of information. Instead, two areas with very diverse service consumption will have a high weighted diversity, and joining them in a same cluster will lose the specificity of the original distributions.

According to ($ii$) above, we can embed weighted diversity as a similarity measure in any clustering algorithm. In this work, we opt for a greedy \emph{divide-et-impera} solution, which, unlike legacy (\eg spectral, agglomerative, modularity-based) clustering techniques, avoids computing pairwise weighted diversities for the billion edges in the complete mesh of geographical areas.
The algorithm, outlined in Alg.\,\ref{alg:clust}, separates the problem in two phases as follows.

\subsubsection{Phase I}

In the first phase, we start by initializing the clustering $\mathfrak{K}$ to the set of communes $\Cm$, \ie considering each commune in a separate cluster (line~\ref{alg:init}). We then compute the dissimilarity matrix $\Dm$ among all communes in $\Cm$ by initializing all values to $\infty$ (line~\ref{alg:infty}), and then updating the actual weighted diversity via the expression in \eqref{eq:div} only between pairs of adjacent communes (lines~\ref{alg:wd-s}-\ref{alg:wd-e}). An important remark is that the matrix $\Dm$ is very sparse, since communes typically have a fairly small number of neighboring areas (\eg less than ten), which is orders of magnitude lower than the cardinality of $\Cm$: therefore, populating $\Dm$ with weighted diversities is dramatically faster than computing the weighted diversity for all pairs of areas in the target region.
At this point, matrix $\Dm$ represents a sparse, weighted graph on which we can run any practical algorithm \texttt{\small cluster} to produce the desired grouping of the original communes in the starting $\mathfrak{K}$ into $N_{K_1}$ clusters (line~\ref{alg:clust1}). In our implementation, we opt for a simple hierarchical clustering technique also presented in Alg.\,\ref{alg:clust}. This is a traditional greedy approach~\cite{banerjee_clust,dhillon_clust2}: until the desired number of clusters is obtained (line~\ref{alg:clust-finish}), it proceeds by identifying the two areas (or clusters) with minimum weighted diversity (line~\ref{alg:min}), removing them from the current set of clusters (line~\ref{alg:remove}), performing a merge of the geographical areas and associated mobile service usage distributions (line~\ref{alg:merge}), and finally adding the new cluster to the updated $\mathfrak{K}$ (line~\ref{alg:add}). The dissimilarity matrix is also updated, by removing the weighted diversity values associated to the two merged areas, and computing and adding to $\Dm$ the weighted diversities between the newly created cluster $J$ and all adjacent (merged) regions in the current $\mathfrak{K}$ (line~\ref{alg:update-s}-\ref{alg:update-e}).

\setlength\floatsep{0pt}
\setlength\textfloatsep{0pt}
\setlength\intextsep{0pt}
\begin{algorithm}[tb]
	\DontPrintSemicolon
	\caption{Two-phase clustering algorithm pseudocode.}
	\label{alg:clust}
	\footnotesize
  \SetKwInput{Input}{input}
	\SetKwFor{Procedure}{procedure}{}{end}
	\SetKwFunction{twoPhaseClustering}{twoPhaseClustering}
	\SetKwFunction{weightedDiversity}{weightedDiversity}
	\SetKwFunction{cluster}{cluster}
	\SetKwFunction{remove}{remove}
	\BlankLine
	\Input{$\Cm$, set of geographical areas, \ie communes}
	\Input{$\Am$, adjacency matrix of areas in $\Cm$}
	\Input{$N_{K_1}$, $N_{K_2}$ target number of clusters in phases I and II}
	\BlankLine
	\Procedure{\twoPhaseClustering{$\Cm,\Am$}}{
		$\mathfrak{K}:\mathcal{C} \rightarrow \mathcal{C}$ s.t. $\mathfrak{K}(c) = c, \forall c \in \mathcal{C}$ \label{alg:init} \;
		$\Dm \in \mathbb{R}^{N_C \times N_C} \leftarrow \infty$ \label{alg:infty} \;
		\ForEach{$(i,i') \in \Am$} { \label{alg:wd-s}
			$\Dm(i,i') =$ \weightedDiversity{$J_\mathfrak{K}(i),J_\mathfrak{K}(i')$} \label{alg:wd} \;
		} \label{alg:wd-e}
		$\mathfrak{K}_1$ $\leftarrow$ \cluster{$\mathfrak{K}, \Dm, \Am, N_{K_1}$} \label{alg:clust1} \;
		$\Dm_1 \in \mathbb{R}^{N_{K_1} \times N_{K_1}} \leftarrow \infty$ \label{alg:second-s} \;
		\ForEach{$(i,i') \in \Dm_1$} {
			$\Dm_1(i,i') =$ \weightedDiversity{$J_{\mathfrak{K}_1}(i),J_{\mathfrak{K}_1}(i')$} \;
		} \label{alg:second-e}
		$\mathfrak{K}_2$ $\leftarrow$ \cluster{$\mathfrak{K}_1, \Dm_1, \Am, N_{K_2}$} \label{alg:second-clust}\;
		\Return{$\mathfrak{K}_2$} \;
	}
	\BlankLine
	\Procedure{\cluster{$\mathfrak{K},\Dm, \Am, N_K$}}{
		\While{$|\mathfrak{K}| > N_K$}{ \label{alg:clust-finish}
			$(i^*,i^{'*}) \leftarrow \argmin_{(i,i')} \Dm(i,i')$ \label{alg:min} \;
			$\mathfrak{K}$ $\leftarrow$ $\mathfrak{K} \setminus J_\mathfrak{K}(i^*),J_\mathfrak{K}(i^{'*})$ \label{alg:remove} \;
			$J$ $\leftarrow$ $J_\mathfrak{K}(i^*) \bigcup J_\mathfrak{K}(i^{'*})$ \label{alg:merge} \;
			$\mathfrak{K}$ $\leftarrow$ $\{\mathfrak{K}, J\}$ \label{alg:add} \;
			\ForEach{$i < |\mathfrak{K}|$}{ \label{alg:update-s}
				\remove{$\Dm, (i,i^*), (i,i^{'*})$} \;
				\If{$\exists (j,j') \in \Am$, s.t. $j\in J_\mathfrak{K}(i), j'\in J$}{
					$\Dm(i,|\mathfrak{K}|) =$ \weightedDiversity{$J_\mathfrak{K}(i),J_\mathfrak{K}(|\mathfrak{K}|)$} \label{alg:update} \;
				}
			} \label{alg:update-e}
		}
		\Return{$\mathfrak{K}$} \;
	}
\end{algorithm}
\setlength\floatsep{12.0pt plus 6.0pt minus 4.0pt}
\setlength\textfloatsep{15.0pt plus 8.0pt minus 5.0pt}
\setlength\intextsep{12.0pt plus 6.0pt minus 4.0pt}

At the end of this phase, a specific clustering $\mathfrak{K}_1$ of $N_{K_1}$ merged communes is selected. Typical approaches for picking $N_{K_1}$ rely on expert knowledge of the system, or stopping rules~\cite{milligan85}. Instead, we opt for a simple and pragmatic strategy. Each iteration of the \texttt{\small cluster} algorithm decreases $N_K$ by generating a refinement of the previous clustering, which, according to Definition 5.1, retains lower or equal information than that available at the previous step. Thus, a larger $N_{K_1}$ is always a better choice, and $N_{K_1}$ can be straightforwardly set to the order of the largest graph that is computationally manageable during the second phase.

\begin{figure}[tb]
	\centering
	\includegraphics[width=0.48\columnwidth]{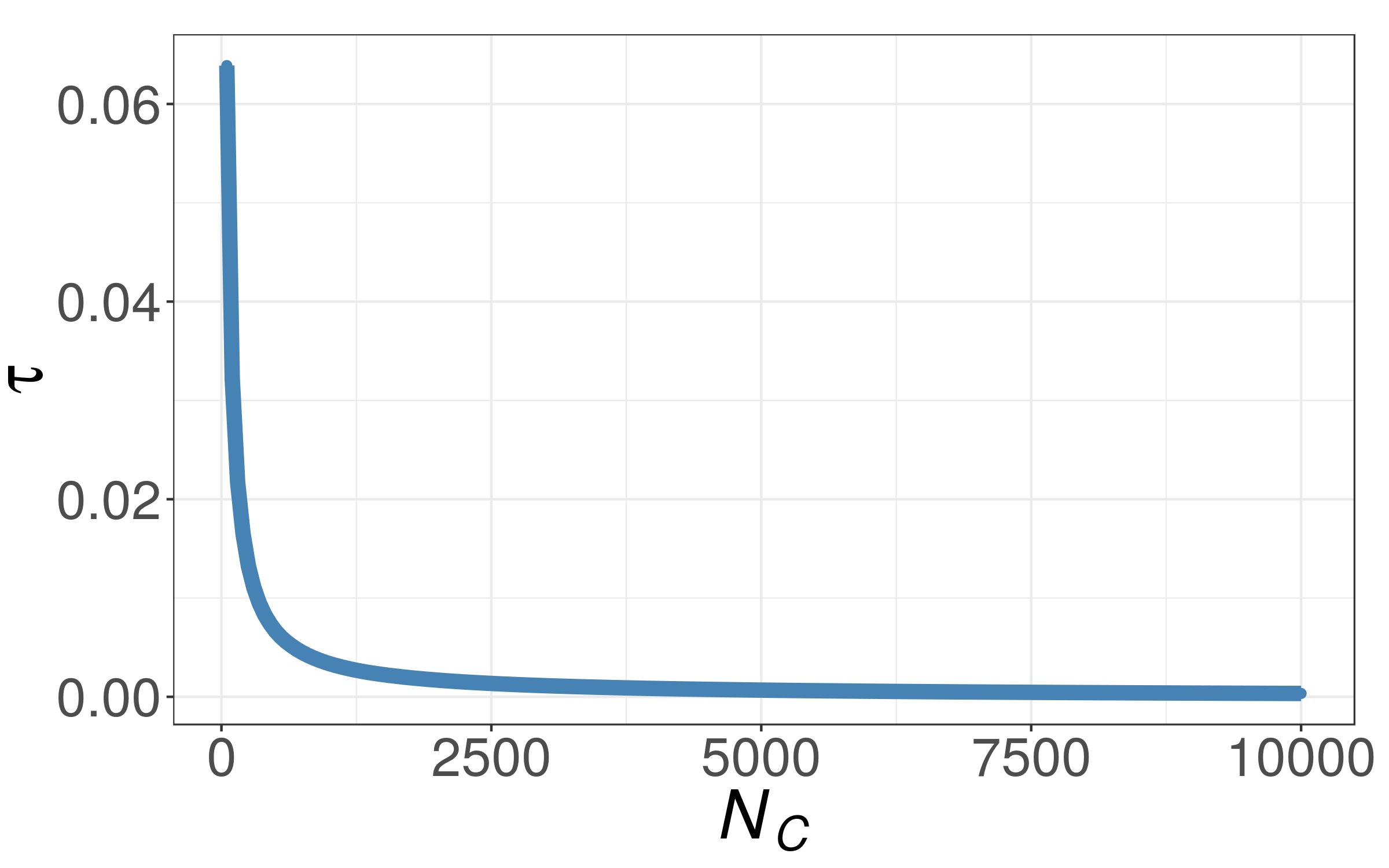}
	\includegraphics[width=0.48\columnwidth]{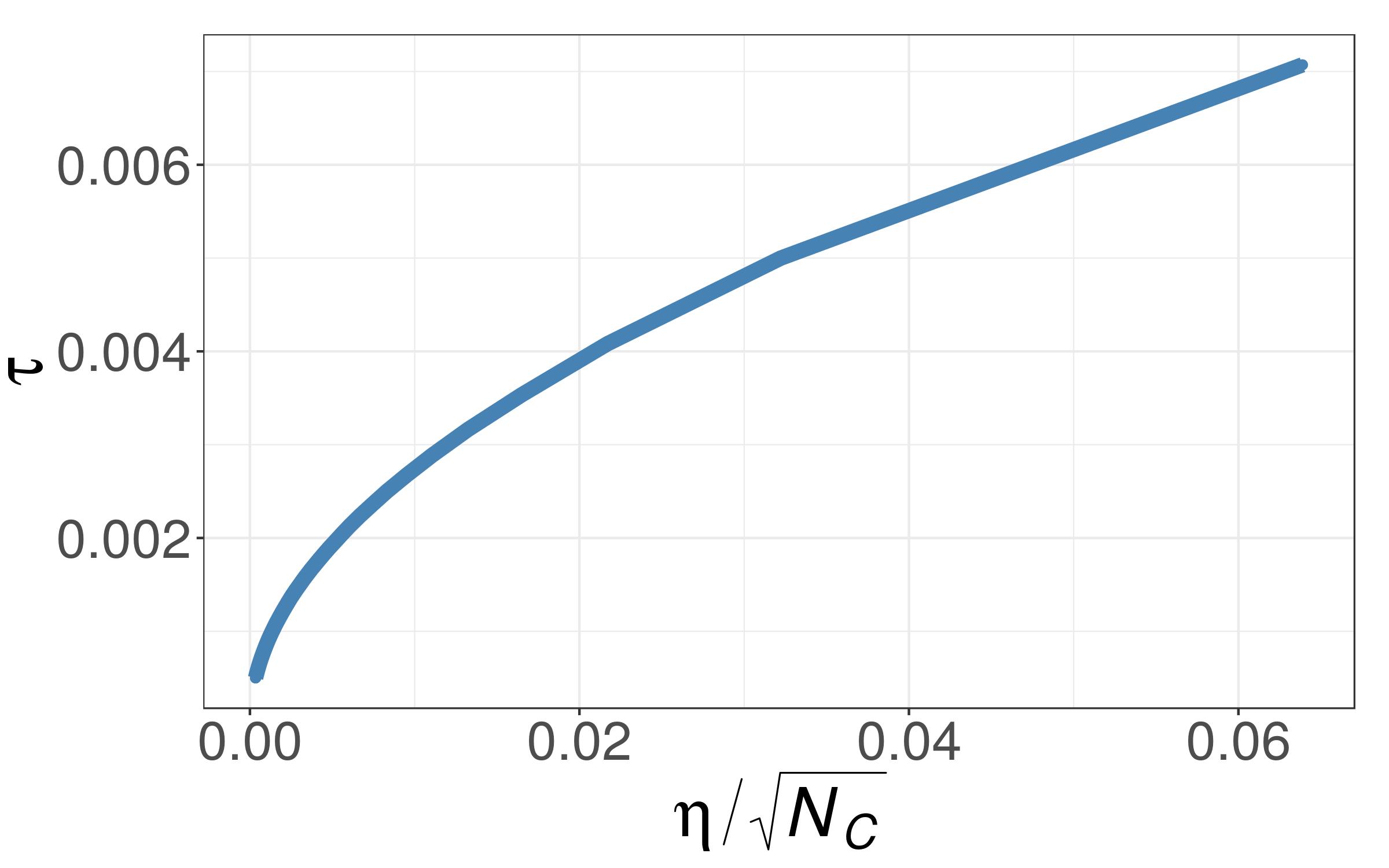}
	\vspace*{-8pt}
	\caption{France scenario. Left: variation of $\tau$ with number of communes. Right: relationship between $\tau$ and $1/\sqrt{N_C}$.}
	\label{fig:tau}
	\vspace*{-15pt}
\end{figure}

\subsubsection{Phase II}

In the second phase, we build a clique, \ie fully connected mesh, of the clusters received from the first phase, and compute the weighted diversity for all pairs of clusters in $\mathfrak{K}_1$ (lines~\ref{alg:second-s}-\ref{alg:second-e}). Note that, unlike in the first phase, this is now a feasible operation, as $N_{K_1}$ is selected by accounting for computational feasibility.

We then run the \texttt{\small cluster} algorithm again on the new graph (line~\ref{alg:second-clust}). This operation allows breaking the spatial proximity constraint imposed during phase I: areas that are geographically distant and yet show similar mobile service distributions can now be grouped together in a scalable way.
Using the weighted diversity as an edge weight metric returns an easily interpretable view of the information loss at each refinement, and allows for an educated choice of the eventual number of clusters to retain, $N_{K_2}$. We provide an example of this property in the France case study, in Section\,\ref{subsec:results}.

\subsubsection{Complexity analysis}

The function \texttt{\small weightedDiversity} operates on the conditional distributions $p_{S|C}(j|i)=\rho_{j,i}$ of $S$ on $C$. Thus, its complexity is linear with respect to the size of the outcome set of $S$, \ie  $\mathcal{O}(N_S)$ for any pair of geographical areas.

During phase I, the calculation of \texttt{\small weightedDiversity} is first repeated for all non-zero elements of the adjacency matrix $\Am\in\mathbb{R}^{N_C \times N_C}$ to obtain $\Dm$, which yields a complexity $\mathcal{O}\left(N_S(\tau N_C)^2\right)$. Here, $\tau$ is the squared fractional average degree in the adjacency graph described by matrix $\Am$: it denotes the sparsity of the matrix due by the geographical topology of the target region. 
Then, each iteration of the \texttt{\small cluster} algorithm requires recomputing all weighted diversities for the neighboring areas of those selected for merging, with complexity $\mathcal{O}\left(N_S (\tau N_C)\right)$. Overall, this leads to a complexity of the first phase $\mathcal{O}\left(N_S(\tau N_C)^2 + N_S (\tau N_C)\right) = \mathcal{O}\left(N_S(\tau N_C)^2\right)$.

In phase II, the same operations above are repeated on a different, fully connected graph with a reduced number of nodes equal to the clusters from the first phase. Hence, the complexity is $\mathcal{O}\left(N_S N_{K_1}^2\right)$.
As $N_{K_1} \ll N_C$, we can approximate the total complexity of the algorithm in the two phases as $\mathcal{O}\left(N_S(\tau N_C)^2\right)$.

A key remark is that, in planar spatial graphs such as those we consider, $\tau$ is typically an extremely low number that scales approximately as $1/\sqrt{N_C}$ when $N_C$ grows~\cite{boccaletti06}. This effectively reduces the complexity of the proposed two-phase algorithm to $\mathcal{O}\left(N_S N_C\right)$, making it extremely efficient in large-scale scenarios. Evidence of the scalability of the approach in is provided in Fig.\,\ref{fig:tau}, for the France nationwide scenario. The left plot shows the evolution of $\tau$ with respect to $N_C$ in the case of the adjacency matrix of communes in France, when considering an increasingly larger portion of the country, \ie higher $N_C$. The curve confirms the scaling property indicated above, which is even more clearly seen in the right plot of Fig.\,\ref{fig:tau}, where $\tau$ is shown to scale as $\eta / \sqrt{N_C}$ with $\eta$ = 0.05.

\subsection{Results in the France case study}
\label{subsec:results}

The two-phase clustering introduced in Section\,\ref{sub:algo} above enables the investigation of mobile service usage across the whole country of France. In the light of the results in Section\,\ref{sub:informative-france}, it makes sense to limit the analysis to the set of informative services $\mathcal{S}'$ that yield non-negligible geographical diversity.

\subsubsection{Phase I analysis}

\begin{figure}[tb]
	\centering
	\subfigure[Mutual Information]{\includegraphics[width=0.48\columnwidth,trim={0 7 0 0},clip]{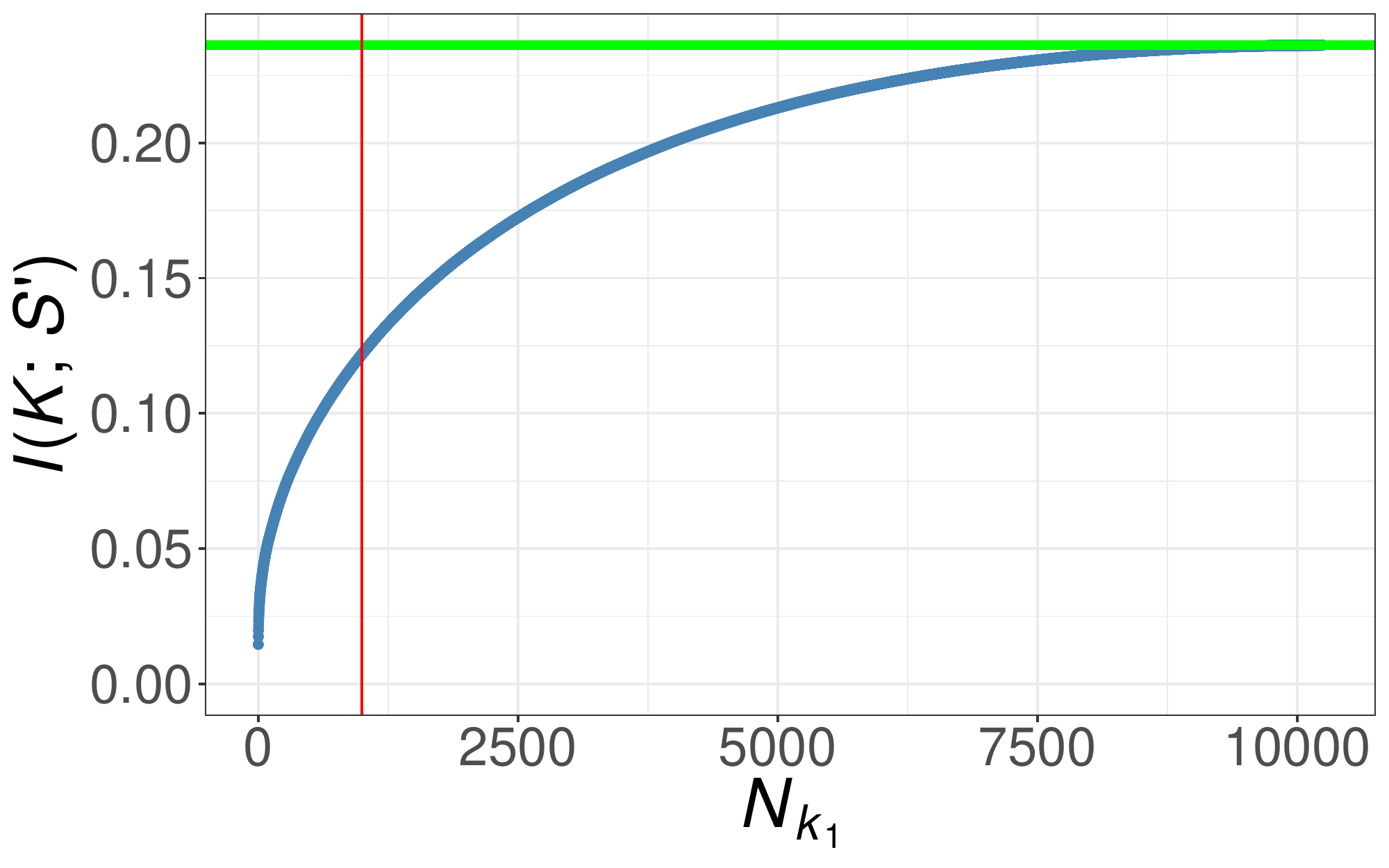}\label{fig:1_mi_cumsum_merged_all_logFALSE_13_communes_10243}}
	\subfigure[Weighted diversity]{\includegraphics[width=0.48\columnwidth,trim={0 7 0 0},clip]{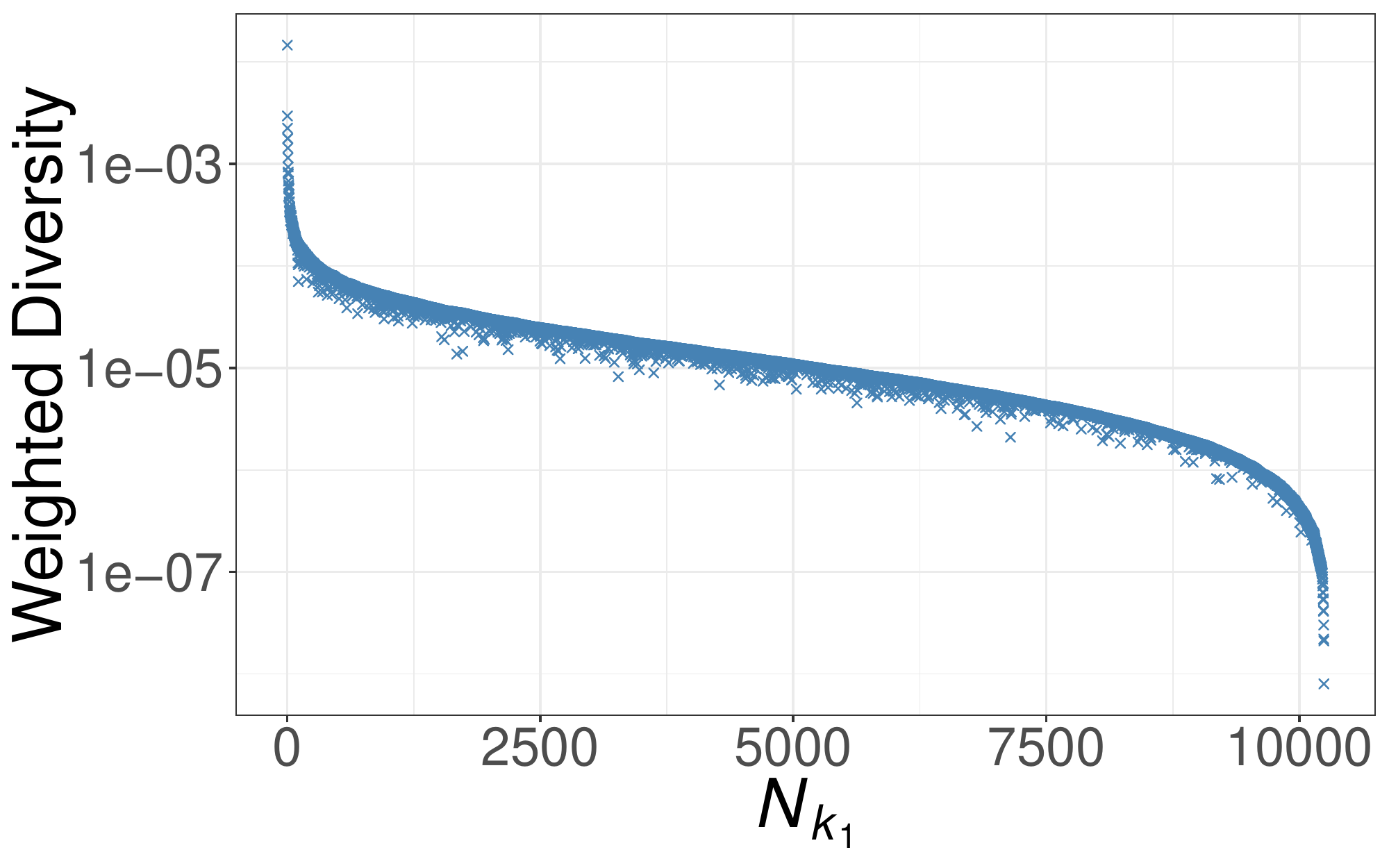}\label{fig:1_mi_perstep_all_logFALSE_13_communes_10243}}
	\vspace*{-12pt}
	\caption{Phase I of the clustering algorithm. Left: Mutual information $I(K;S')$ retained by the clustering versus the number of returned clusters $N_{K_1}$. Right: weighted diversity of the two communes (or clusters) aggregated at each step.}
	\vspace*{-13pt}
	\label{fig:phase1_MI}
\end{figure}

Fig.\,\ref{fig:phase1_MI} shows the mutual information $I(K;S')$ retained by all clusterings $\mathfrak{K}_1$ in the hierarchical structure formed by the greedy approach during the first phase of the algorithm. Looking at each of these plots from right to left allows imagining how the algorithm works. Specifically, $I(K;S')$ is illustrated as a function of $N_{K_1}$, \ie the number of clusters. In the left plot, we note that $I(K;S')$ grows with $N_{K_1}$, as expected from Definition 5.1, until it reaches the complete mutual information $I(C;S')$ (green horizontal line) that characterizes the system of $N_{S'}$ (\ie 13) services and $N_C$ (\ie 10,000) communes in France, when $N_{K_1}=N_C$.

The curve grows faster at first, to slow down afterwards: this means that the majority of the information is retained by the very first clusters, or, equivalently, that the latest iterations of the agglomerative clustering are those that lose the highest informations. This is highlighted in the right plot of Fig.\,\ref{fig:phase1_MI}, where the merging of the few tens of clusters (on the left) results in an information loss, quantified by the weighted diversity, which is orders of magnitude higher than that incurred at earlier stages of the algorithm (to the right). The diversity gently degrades across the vast majority of $N_{K_1}$ values, with the exception of the last clusterings: this implies that only the opening aggregations do not lose information.

\subsubsection{Phase II analysis}

\begin{figure}[tb]
	\centering
	\subfigure[Mutual Information]{\includegraphics[width=0.48\columnwidth,trim={0 7 0 0},clip]{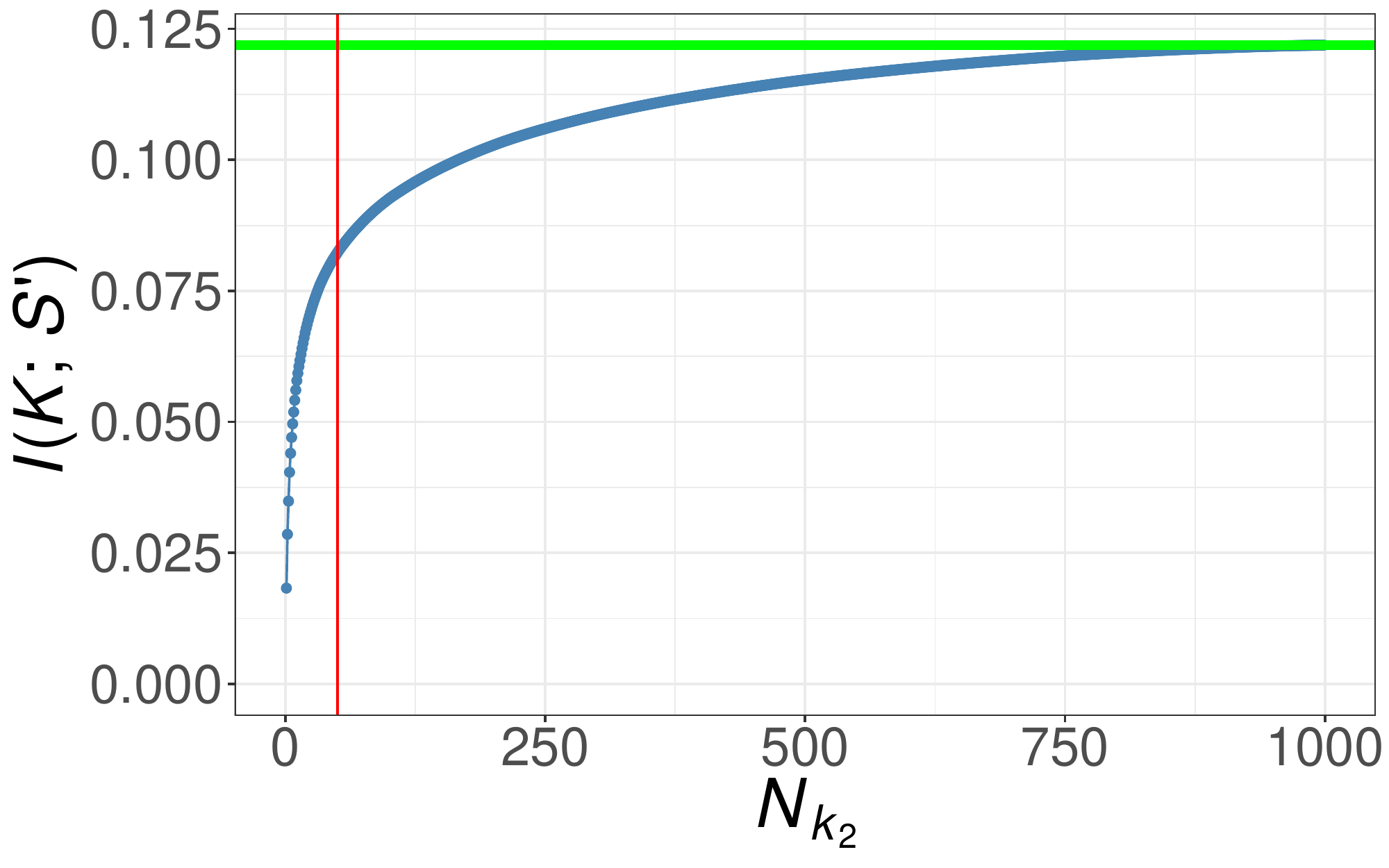}\label{fig:1_mi_cumsum_merged_all_logFALSE_13_communes_999}}
	\subfigure[Weighted diversity]{\includegraphics[width=0.48\columnwidth,trim={0 7 0 0},clip]{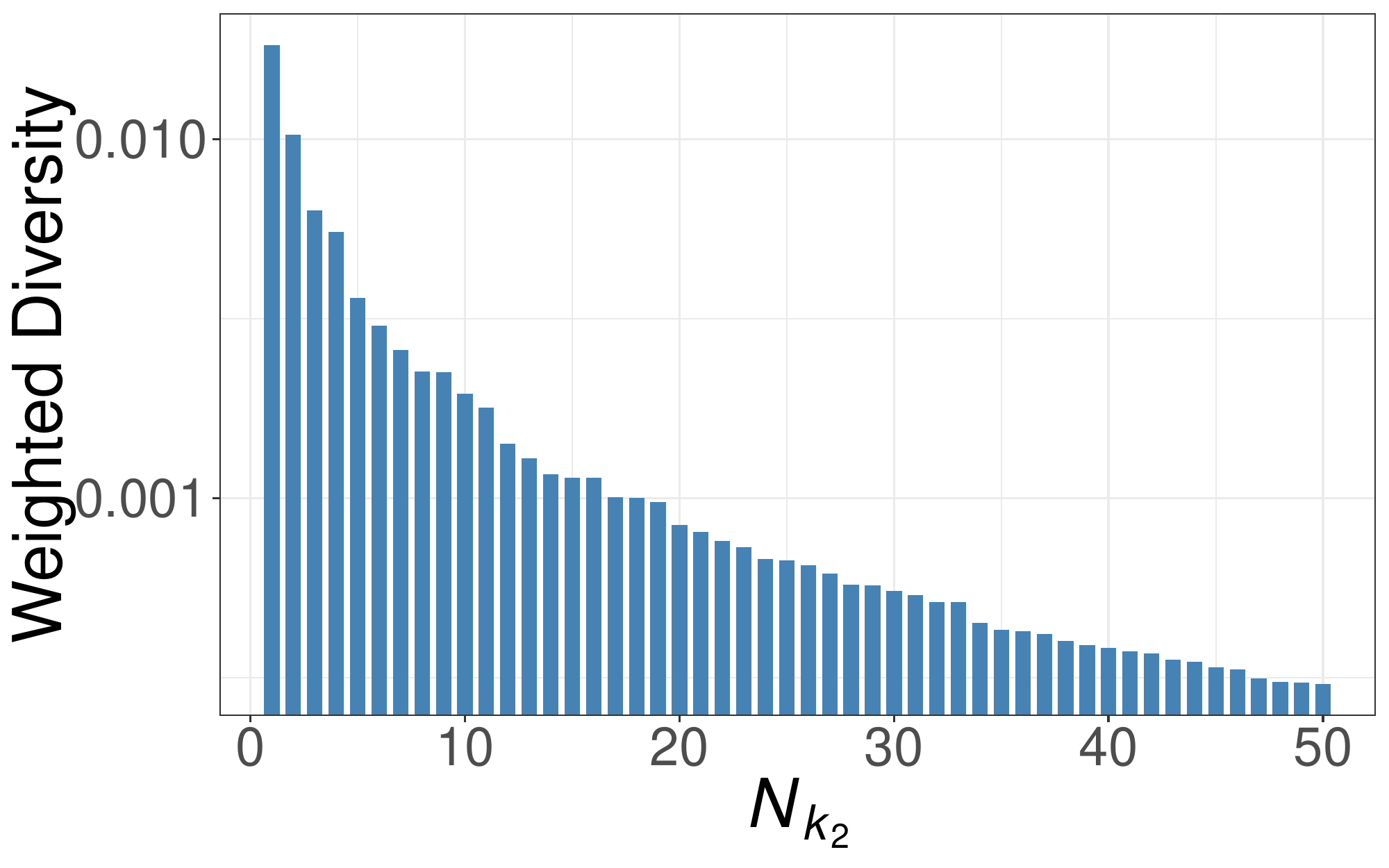}\label{fig:1_mi_perstep_all_logFALSE_13_communes_999}}
	\vspace*{-12pt}
	\caption{Phase II of the clustering algorithm. Left: Mutual information $I(K;S')$ retained by the clustering versus the number of returned clusters $N_{K_2}$. Right: weighted diversity of the two communes (or clusters) aggregated at each step.}
	\vspace*{-6pt}
	\label{fig:phase2_MI}
\end{figure}

As recommended in Section\,\ref{sub:algo}, we retain the maximum number of clusters for which we can afford the computationally expensive processing of the second phase.
We thus consider $N_{K_1}=1,000$, which is highlighted by the vertical line in Fig.\,\ref{fig:phase1_MI}, and captures $52\%$ of the original mutual information. This means that we are bounding the computational complexity of the analysis in phase II to operations on graphs with $500,000$ edges.

Equivalent curves to those for the first phase, illustrating the output of the clustering on the fully connected mesh of 1,000 clusters of communes from the first phase, are shown in Fig.\,\ref{fig:phase2_MI}. Interestingly, the mutual information curve, in the left plot, grows much faster with $N_{K_2}$ than it did with $N_{K_1}$ during phase I. We ascribe this phenomenon to the fact that the clique representation removes all geographical constraints, and grants higher flexibility during the clustering process, allowing matching and merging (clusters of) communes that are spatially distant but showing fairly correlated mobile service distributions. Therefore, the result implicitly proves that similar mobile application usages often occur in areas located at significant geographical distance in France.

The sudden increase of the mutual information also lets us take an easy, informed choice about a reasonable number of clusters: \eg by selecting $N_{K_2}=50$ (the red vertical line in Fig.\,\ref{fig:phase2_MI}), it is possible to retain $67.5\%$ of the mutual information of the 1,000 clusters considered after first phase, and $35\%$ of the total mutual information of the system. Or, a very small $N_{K_2}=9$ preserves $44\%$ and $23\%$ of the mutual information in the two cases. The detailed weighted diversity values are in the right plot of Fig.\,\ref{fig:phase2_MI}.

Overall, the observations above let us formulate our third takeaway message: \textit{\textbf{a small number of commune clusters (\eg 9, equal to a fraction 0.0009 of around 10,000 communes considered in the analysis) is sufficient to retain a substantial percentage (\eg over 20\%) of the diversity observed in the usage of mobile services in France}}. These figures refer to the set $\mathcal{S}'$ of services that yield the highest spatial heterogeneity. Then, the takeaway above implies that there exists a fairly limited set of typical behaviors in the usage of mobile services across the whole country, even when considering only those applications that show substantial geographic diversity.

\begin{figure*}[tb]
	\centering
	\includegraphics[height=0.53\columnwidth]{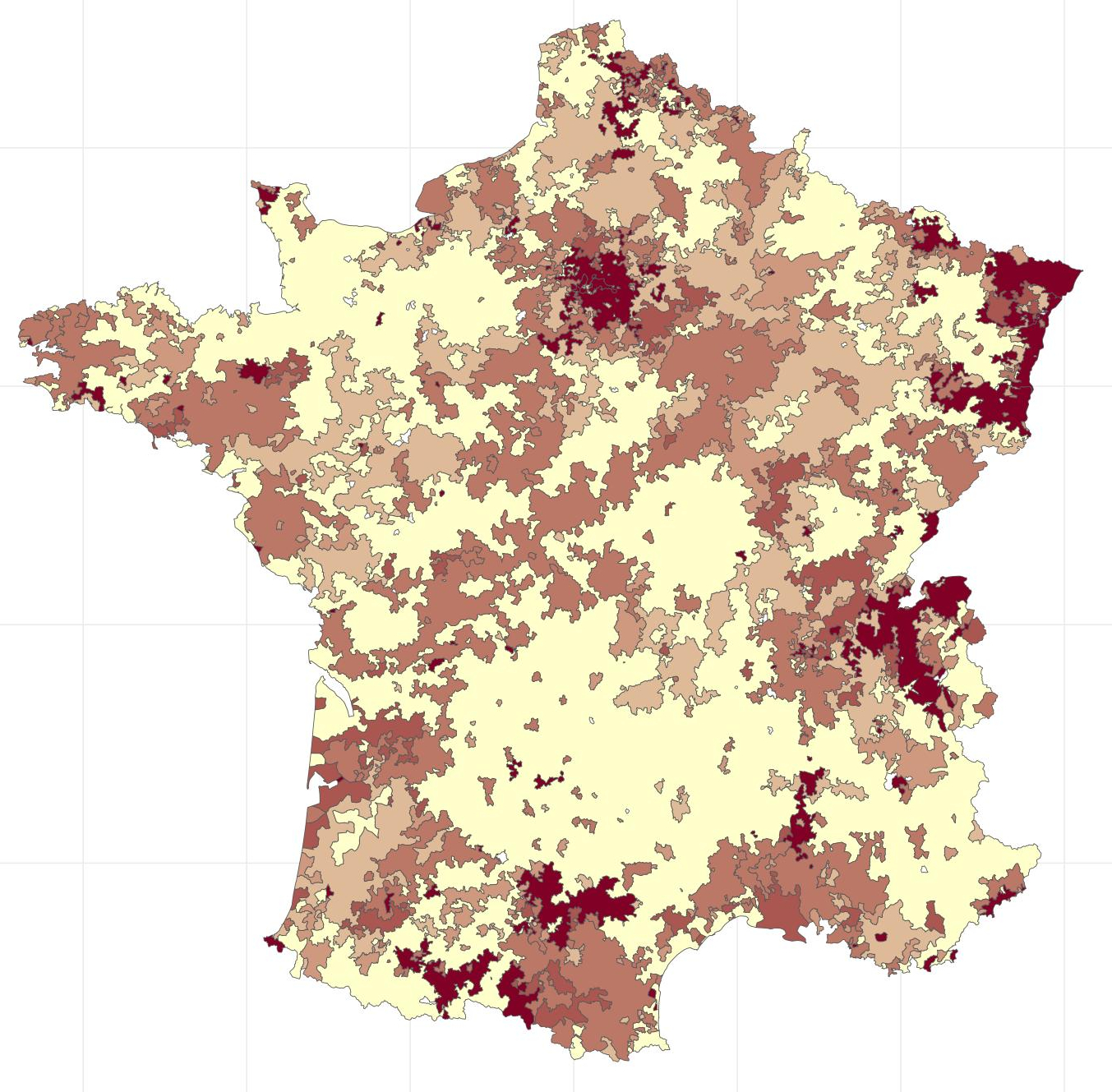}
	\hspace*{12pt}
	\includegraphics[height=0.53\columnwidth]{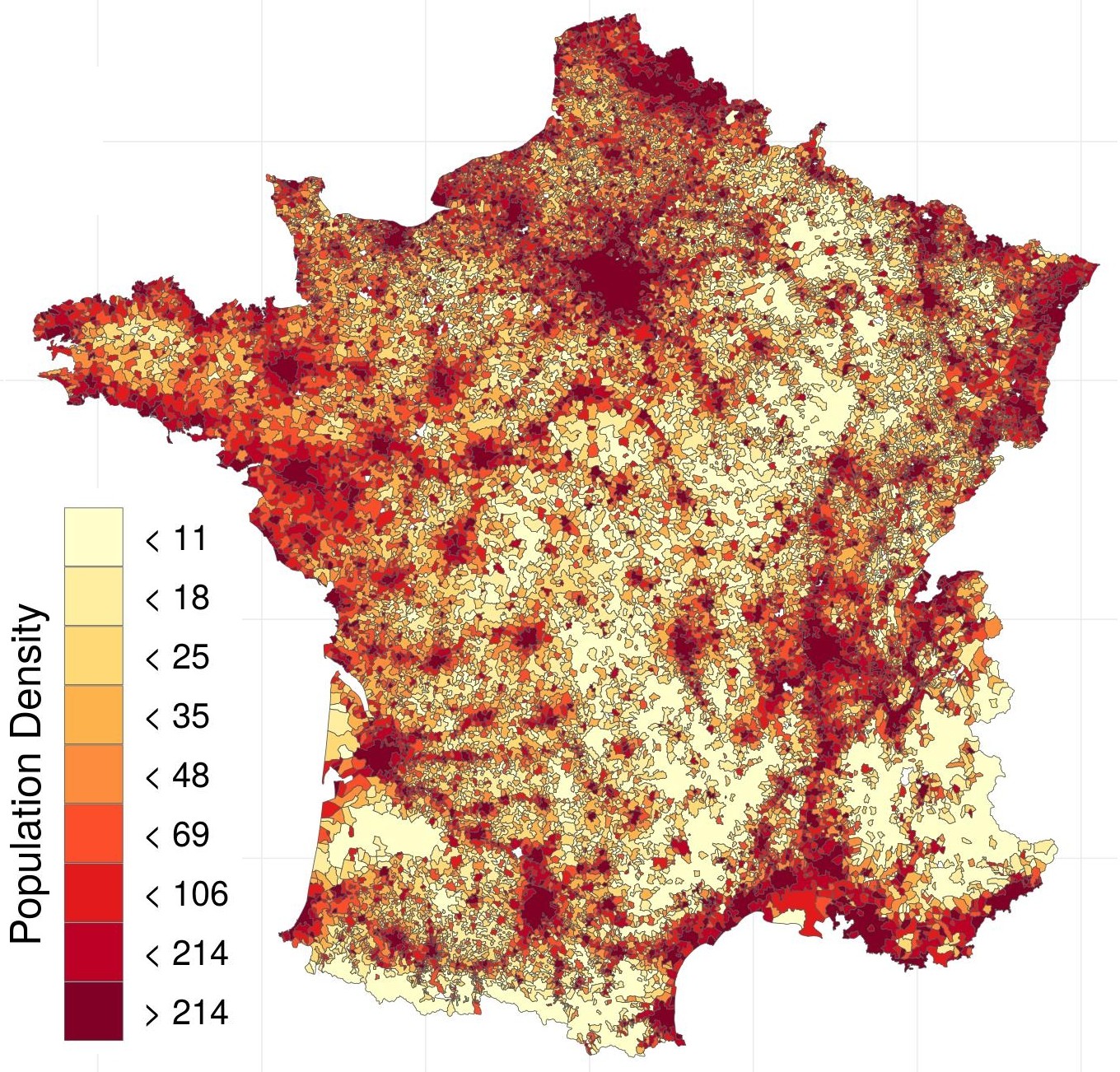}
	\hspace*{12pt}
	\includegraphics[height=0.54\columnwidth]{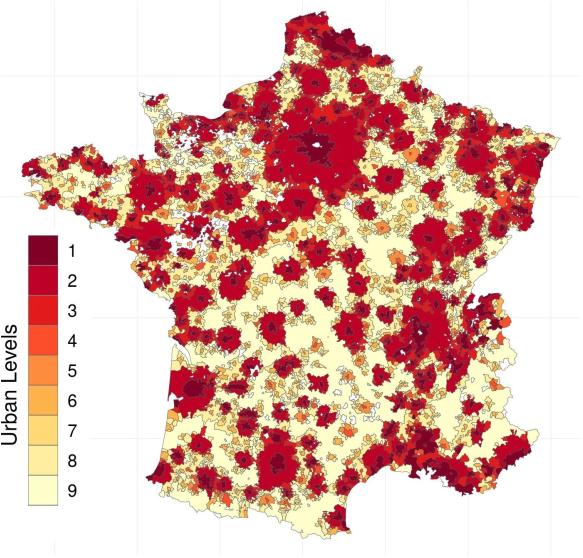}
	\vspace*{-4pt}
	\caption{Maps of France. Left: geographical layout of the $N_{K_2}=9$ clusters produced by our algorithm, each identified by a color. Middle: population density levels (in people per square kilometer). Right: urbanization levels. Figure best viewed in colors.}
	\label{fig:map9}
	\vspace*{-10pt}
\end{figure*}

\begin{table}[tb]
	\centering
	\caption{Urbanization levels categorizing French communes. Metropolis, medium- and small-sized cities are based on the number of workplaces. Intermediate classes are obtained by geographical adjacency to the three types of urban areas.}
	\vspace*{-8pt}
	\fontsize{7}{2}\selectfont%
	\renewcommand{\arraystretch}{4}
	\setlength{\tabcolsep}{6pt}
	\begin{tabular}{|r|l|r|r|}
		\hline
		{\it Level} & {\it Urbanization class} & {\it Workplaces} & {\it Adjacency} \\
		\hline
		\hline
		1 & Large metropolis & 10,000 and more & - \\
		2 & Large metropolis suburbs & - & to 1 \\
		3 & Large metropolis influence area & - & to 2 \\
		4 & Medium-sized city & 5,000 - 10,000 & - \\
		5 & Medium-sized city suburbs & - & to 4 \\
		6 & Small-sized city & 1,500 - 5,000 & - \\
		7 & Small-sized city suburbs & - & to 6 \\
		8 & Town & - & to 7 \\
		9 & Rural area & - & - \\
		\hline
	\end{tabular}
	\label{tab:urblev}
	\vspace*{-8pt}
\end{table}

\begin{figure*}[tb]
	\centering
	\subfigure[Cluster $k=2$]{\label{fig:}
		\includegraphics[height=0.35\columnwidth,trim={0 0 0 0},clip]{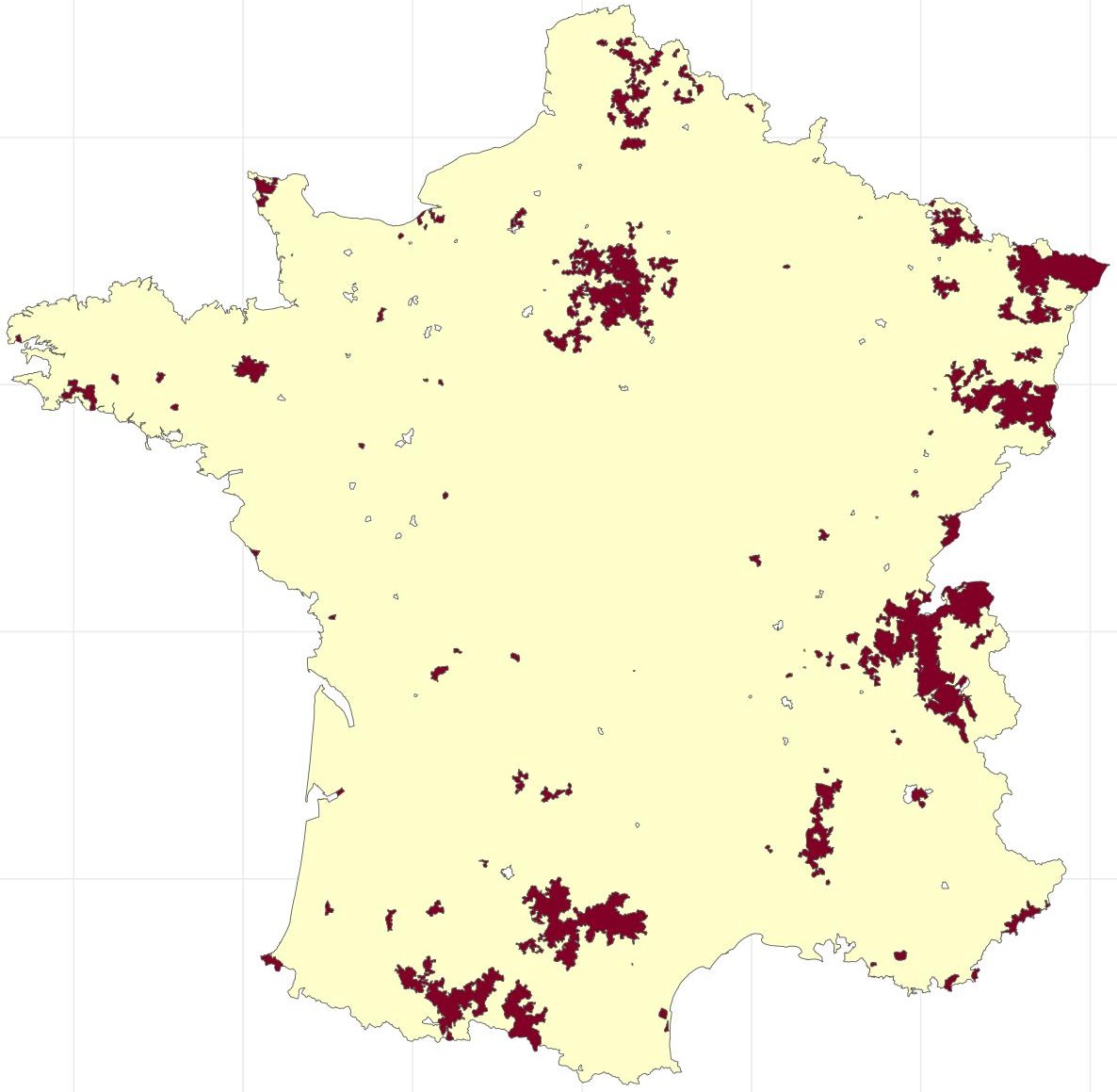}
		\includegraphics[height=0.40\columnwidth,trim={0 0 0 0},clip]{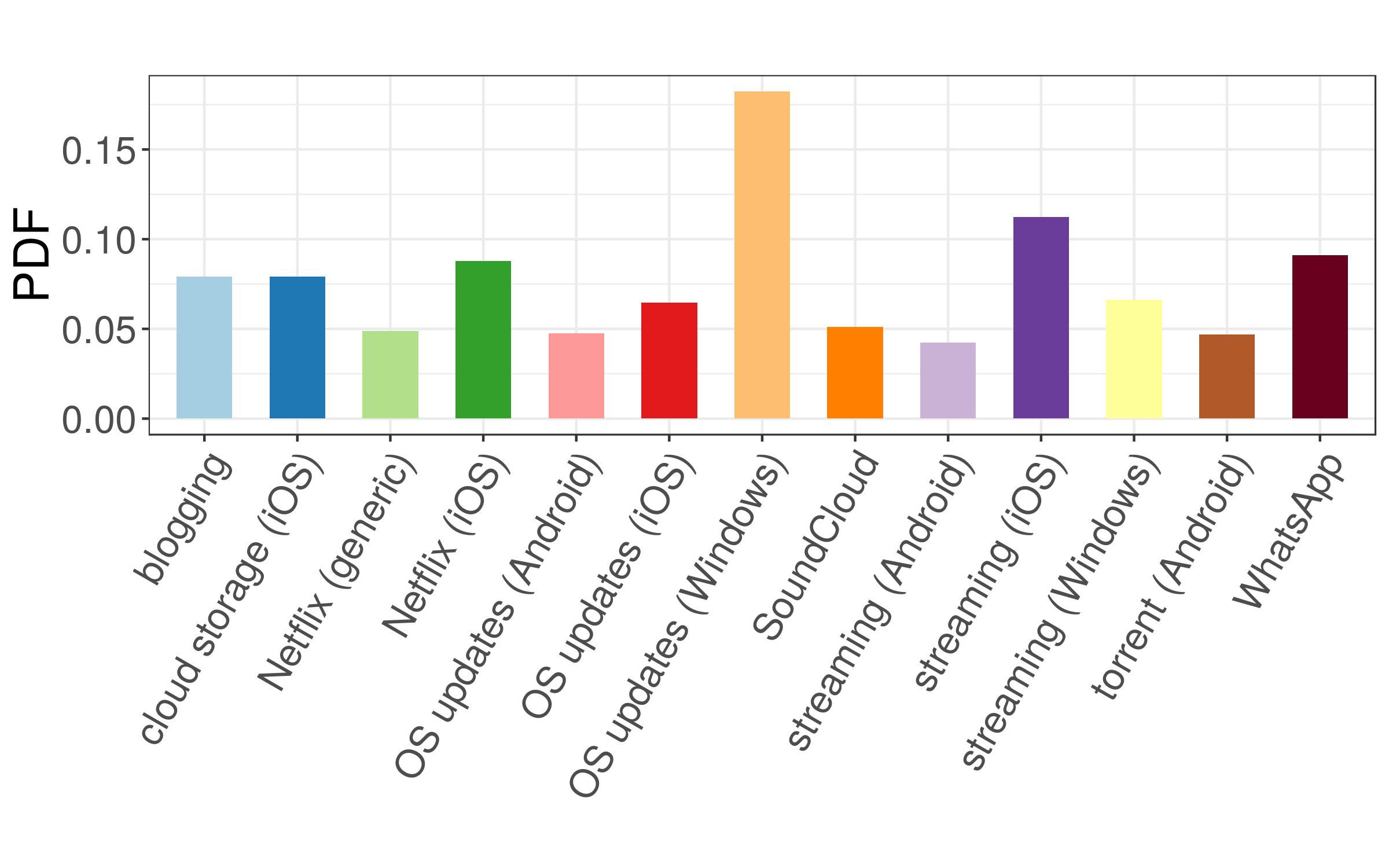}
	}
	\subfigure[Cluster $k=3$]{\label{fig:}
		\includegraphics[height=0.35\columnwidth,trim={0 0 0 0},clip]{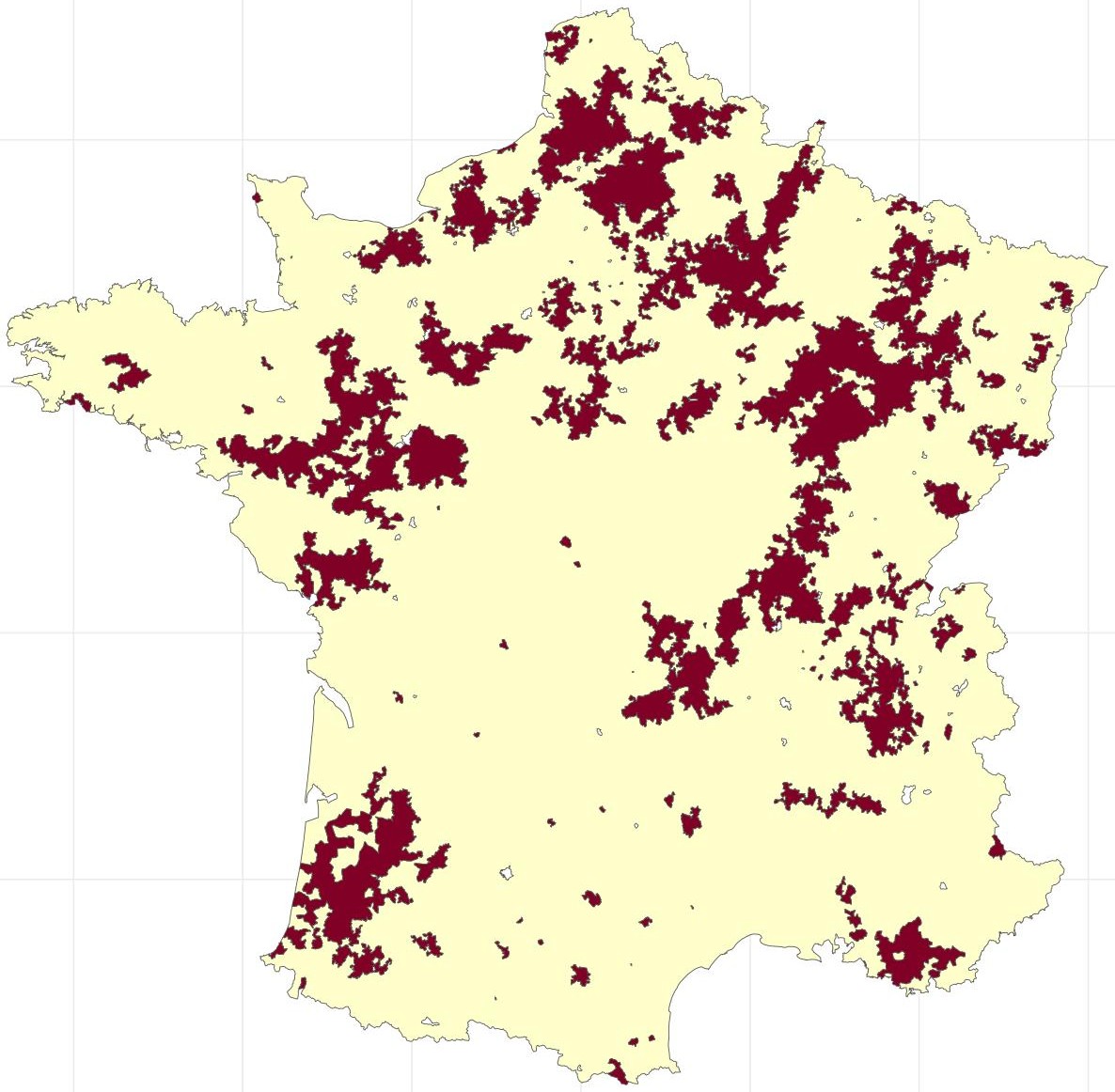}
		\includegraphics[height=0.40\columnwidth,trim={0 0 0 0},clip]{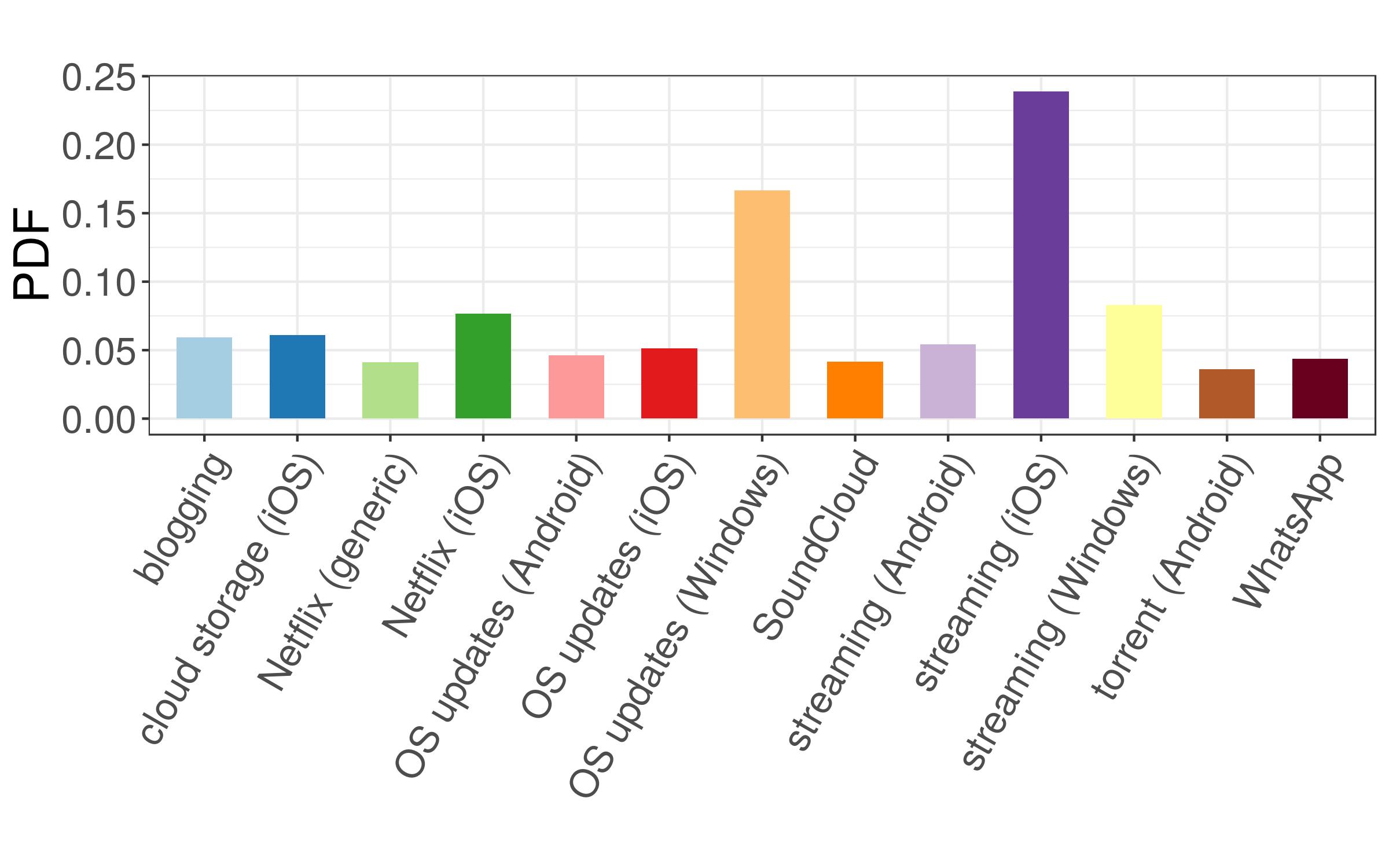}
	}\\
	\vspace*{-8pt}
	\subfigure[Cluster $k=5$]{\label{fig:}
		\includegraphics[height=0.35\columnwidth,trim={0 0 0 0},clip]{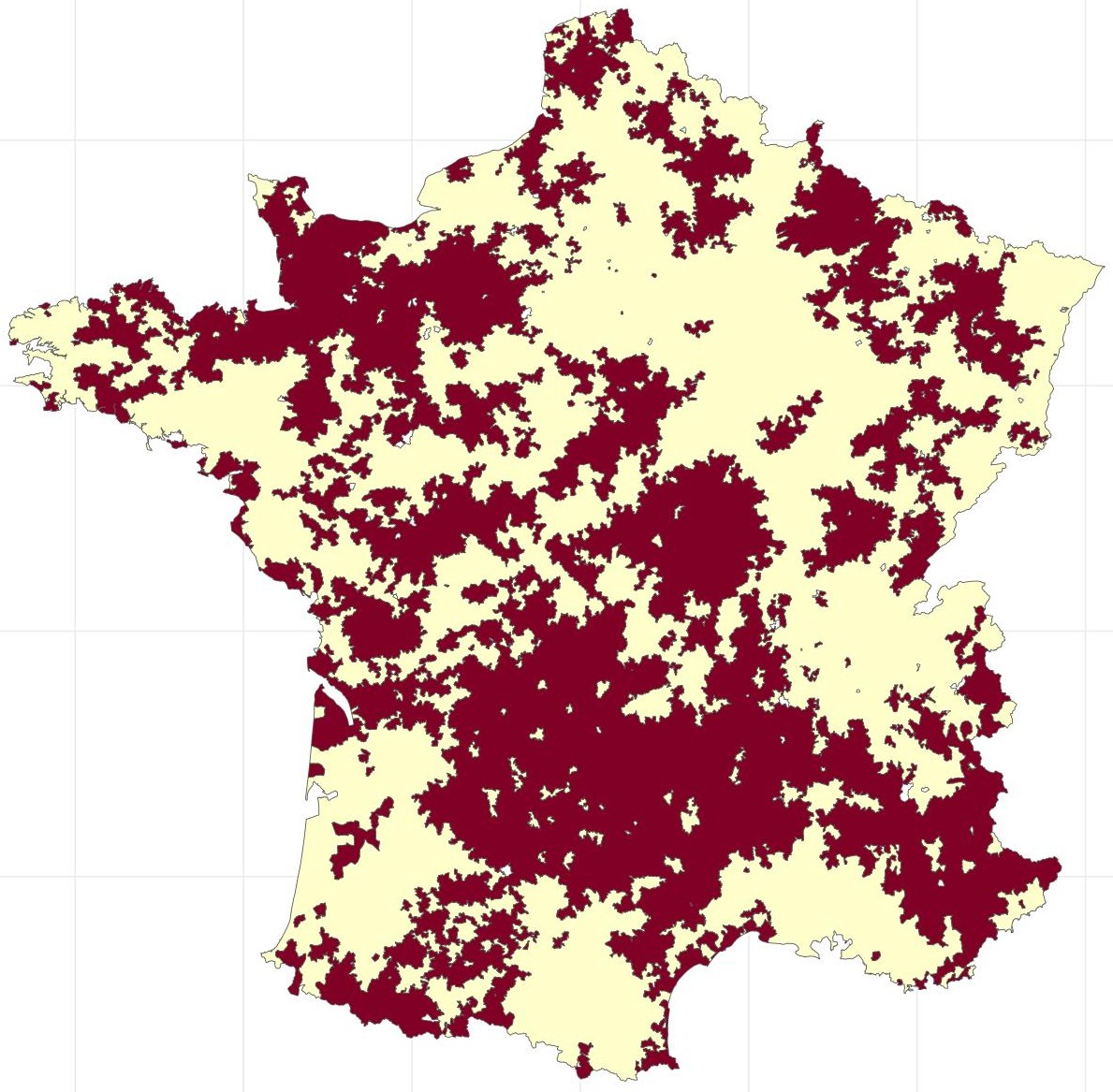}
		\includegraphics[height=0.4\columnwidth,trim={0 0 0 0},clip]{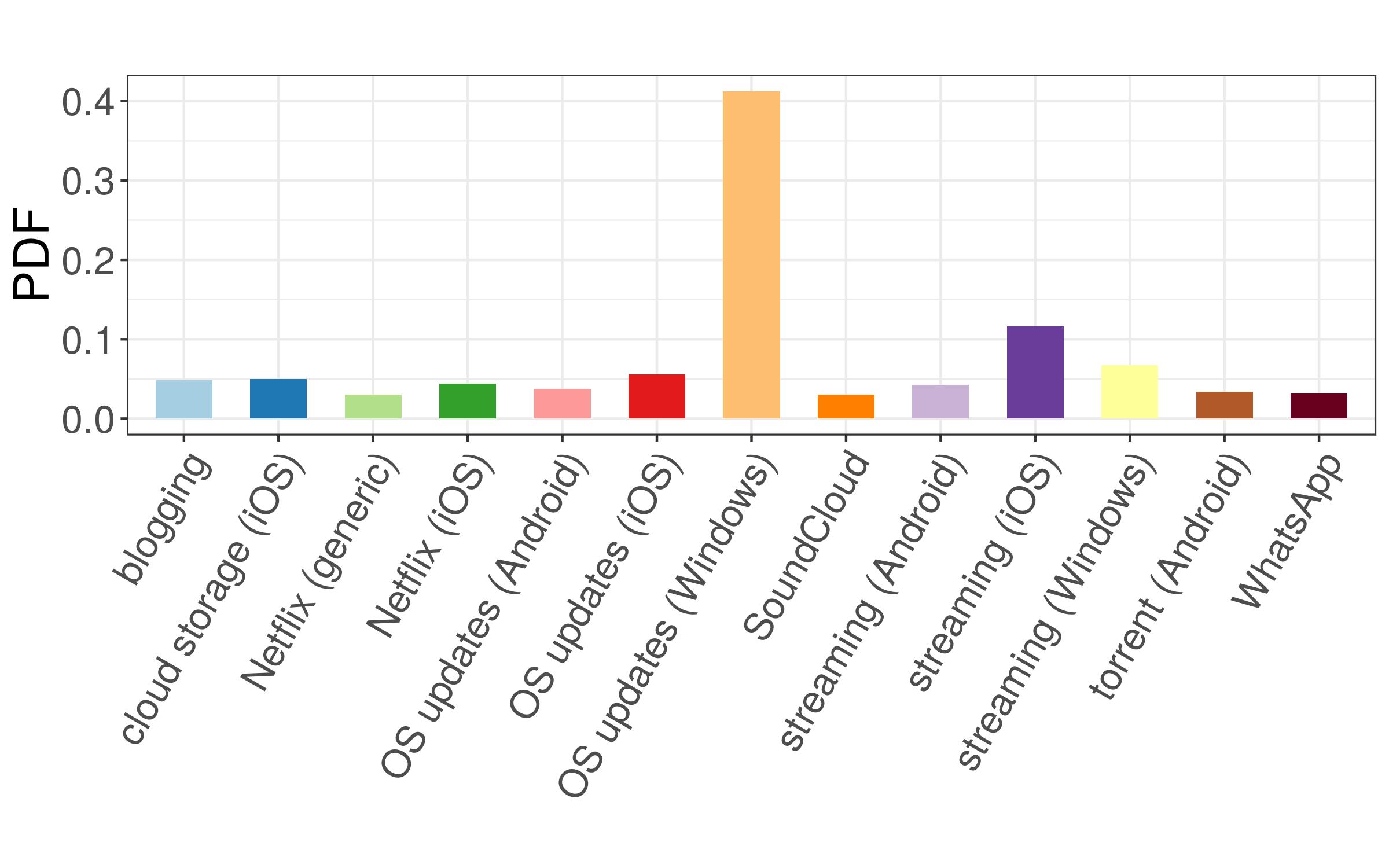}
	}
	\subfigure[Cluster $k=9$]{\label{fig:}
		\includegraphics[height=0.35\columnwidth,trim={0 0 0 0},clip]{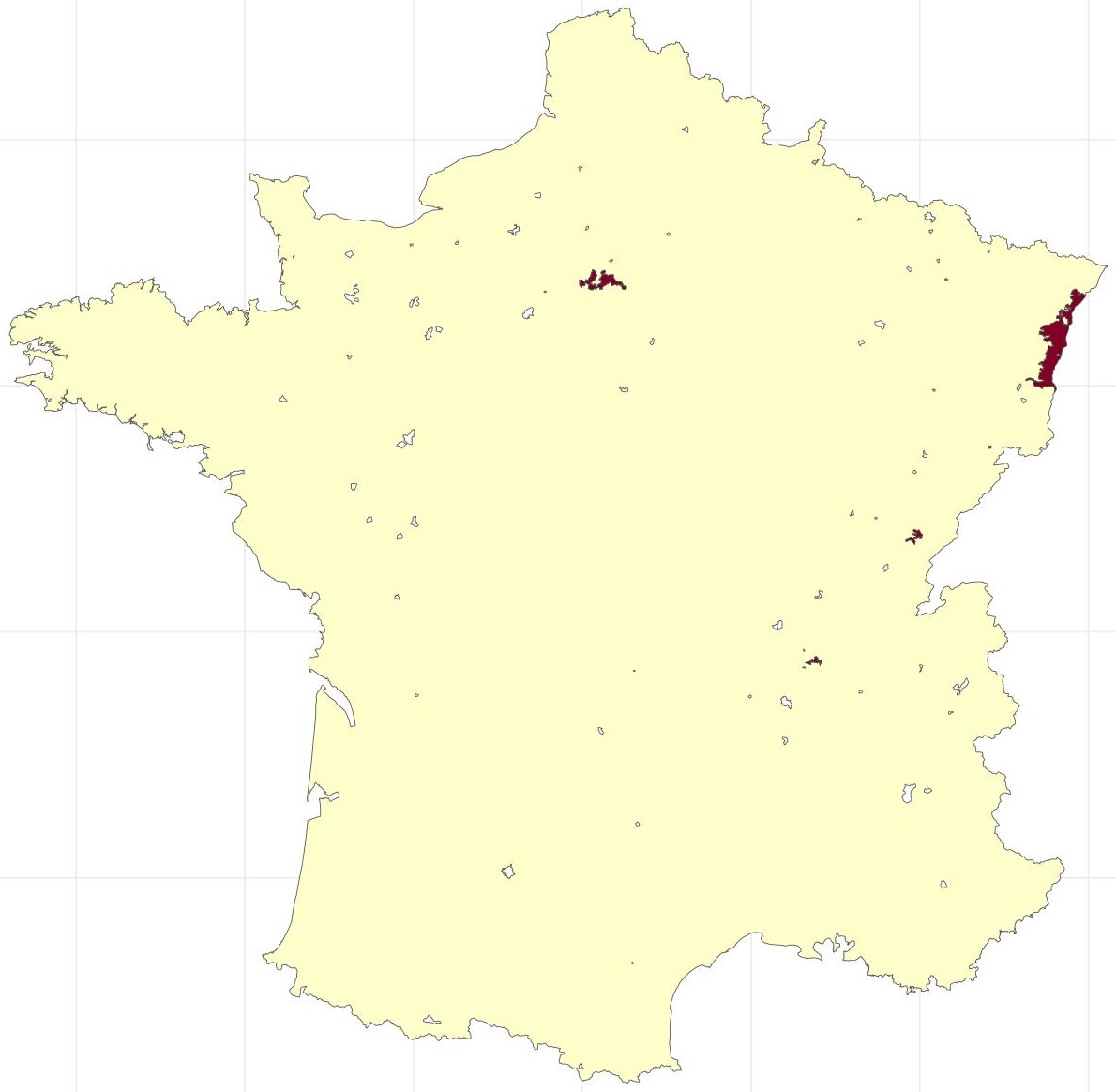}
		\includegraphics[height=0.40\columnwidth,trim={0 0 0 0},clip]{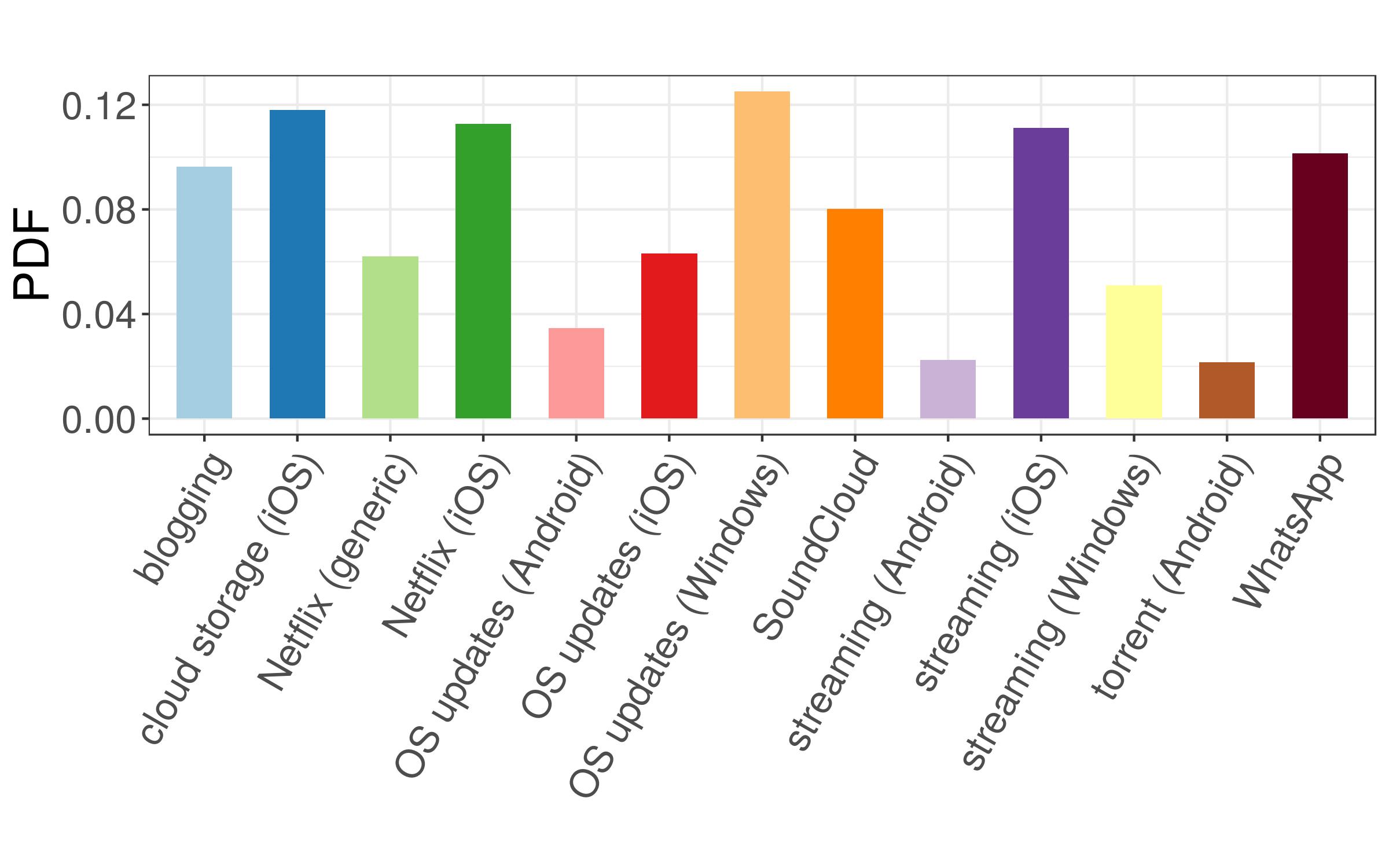}
	}
	\vspace*{-8pt}
	\caption{Geographical coverage of four representative clusters, and associated PDF $\rho_{j,k}$ of the informative service demands.}
	\label{fig:pdf}
	\vspace*{-6pt}
\end{figure*}

\section{Interpretation of results}
\label{sec:interpretation}

In order to better understand the results obtained so far, we investigate the links between the clustered mobile service usages and the country demographics in the representative case of $N_{K_2}=9$.

\subsection{French geography of mobile service usage}

The left plot in Fig.\,\ref{fig:map9} provides an illustration of how clusters are associated to different geographical areas in France. The spatial pattern of clusters is not random. For instance, one cluster clearly tells apart the Paris metropolitan area from its surroundings (dark patch at the center top of the map); or, most of the areas in central France are clustered together in a large continuous region (wide light territory at center bottom of the map).

In fact, a simple visual inspection reveals that the map of clusters yields substantial resemblances to those of two important demographic measures: ($i$) population density, \ie the number of dwelling units per commune, measured in inhabitants/Km\textsuperscript{2}; ($ii$) the urbanization level, \ie a categorization of communes based on number of workplaces and mutual geographical adjacency, which results in the nine categories in Tab.\,\ref{tab:urblev} \cite{insee}. The likeness is evident when comparing the left plot in Fig.\,\ref{fig:map9} with the middle and right plots in the same figure, which respectively portray the population density and urbanization levels for all communes in France.

Also, it is interesting to observe how the most informative services are consumed in these clusters. Fig.\,\ref{fig:pdf} shows the Probability Density Function (PDF) $p_{S'|K}(j|k)=\rho_{j,k}$ of services $j\in\mathcal{S}'$ in a selected subset of clusters $k\in\mathcal{K}$. For the sake of clarity, the PDFs are accompanied by maps displaying the regions included in each cluster. We observe that the distributions yield significant differences, and characterize very heterogeneous surfaces. For instance, clusters 3 and 5 cover large regions all over France that appear to be fairly complementary; the former happens to be characterized by a distinctively high usage of streaming services, while the latter has a high incidence of background traffic generated by smartphones that run Windows Mobile as the operating system.
Instead, clusters 2 and 9 cover much smaller areas in France, and have more balanced distributions across all services.

\subsection{Linking services and demographics}
\label{subsec:linking}

The qualitative analysis above suggests: ($i$) an apparent correlation between the geographical layout of the clusters and the intensity of human presence (captured by the population density and urbanization level); and ($ii$) striking differences in the way specific services are consumed across clusters. By combining these observations, the identified clusters allow bonding mobile services to demographics.

\subsubsection{Methodology}

Let $\mathcal{D}=\{1,\dots,N_D\}$ be a set of demographics levels (either discretized population density levels, or urbanization levels), with cardinality $N_D$. All geographical areas $i\in\mathcal{C}$ are assigned a unique level in $d\in\mathcal{D}$. We can then define as $D$ the random variable with outcome in $\mathcal{D}$ that denotes the probability that a given area $i$ is characterized by a specific demographics level. The joint distribution of demographics levels and clusters is $p_{D,K}(d,k)$, and we can compute conditional probabilities $p_{D|K}(d|k)$ that describe the distribution of individual demographics levels within cluster $k$.

\begin{figure*}[tb]
	\centering
	\subfigure[Cluster 2]{\includegraphics[width=0.52\columnwidth,trim={0 25 0 0},clip]{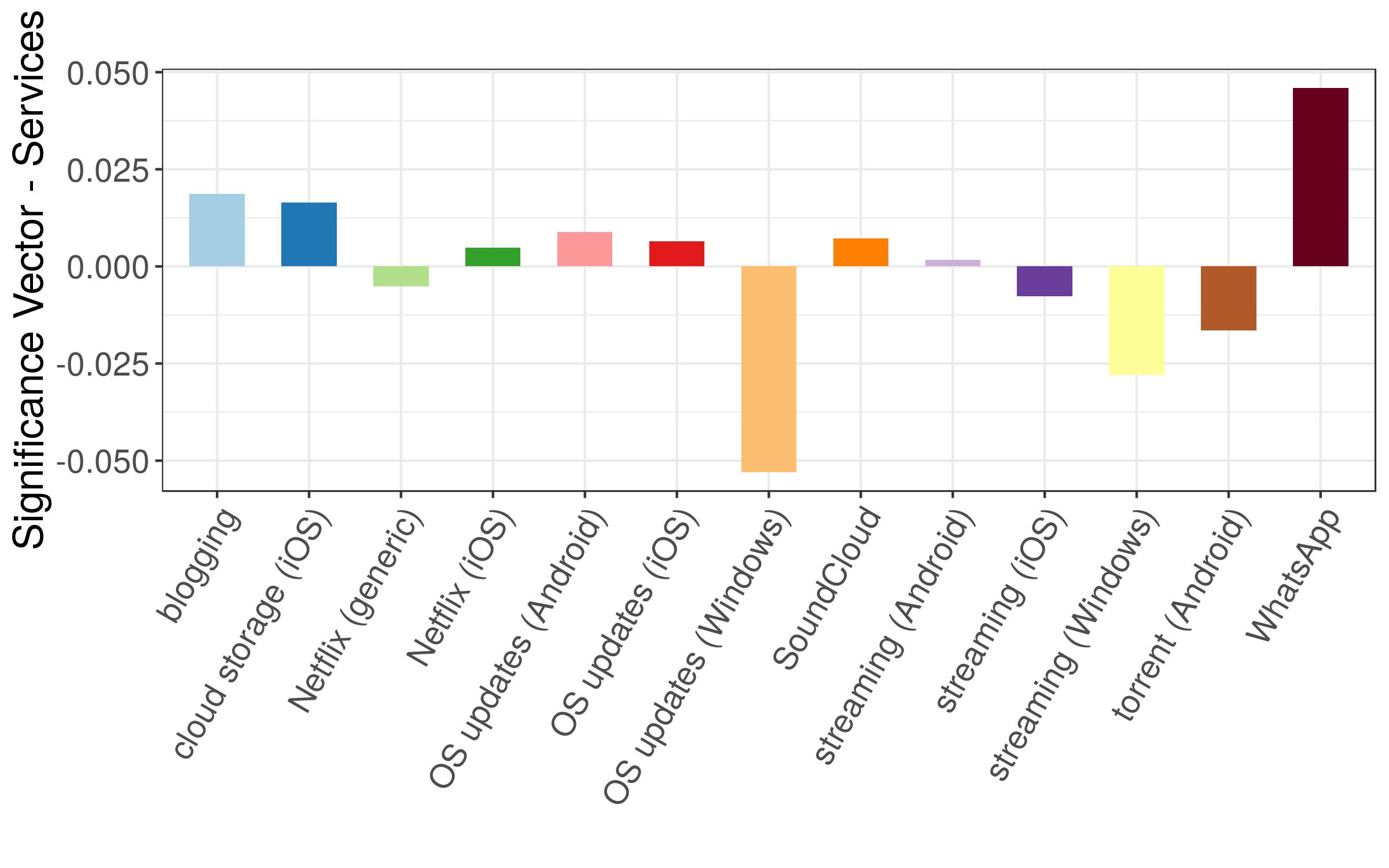}\label{fig:1_PDF_clustered_difference_step_9_cl_2_logFALSE_13_communes_999}}
	\hspace*{-2mm}
	\subfigure[Cluster 3]{\includegraphics[width=0.52\columnwidth,trim={0 25 0 0},clip]{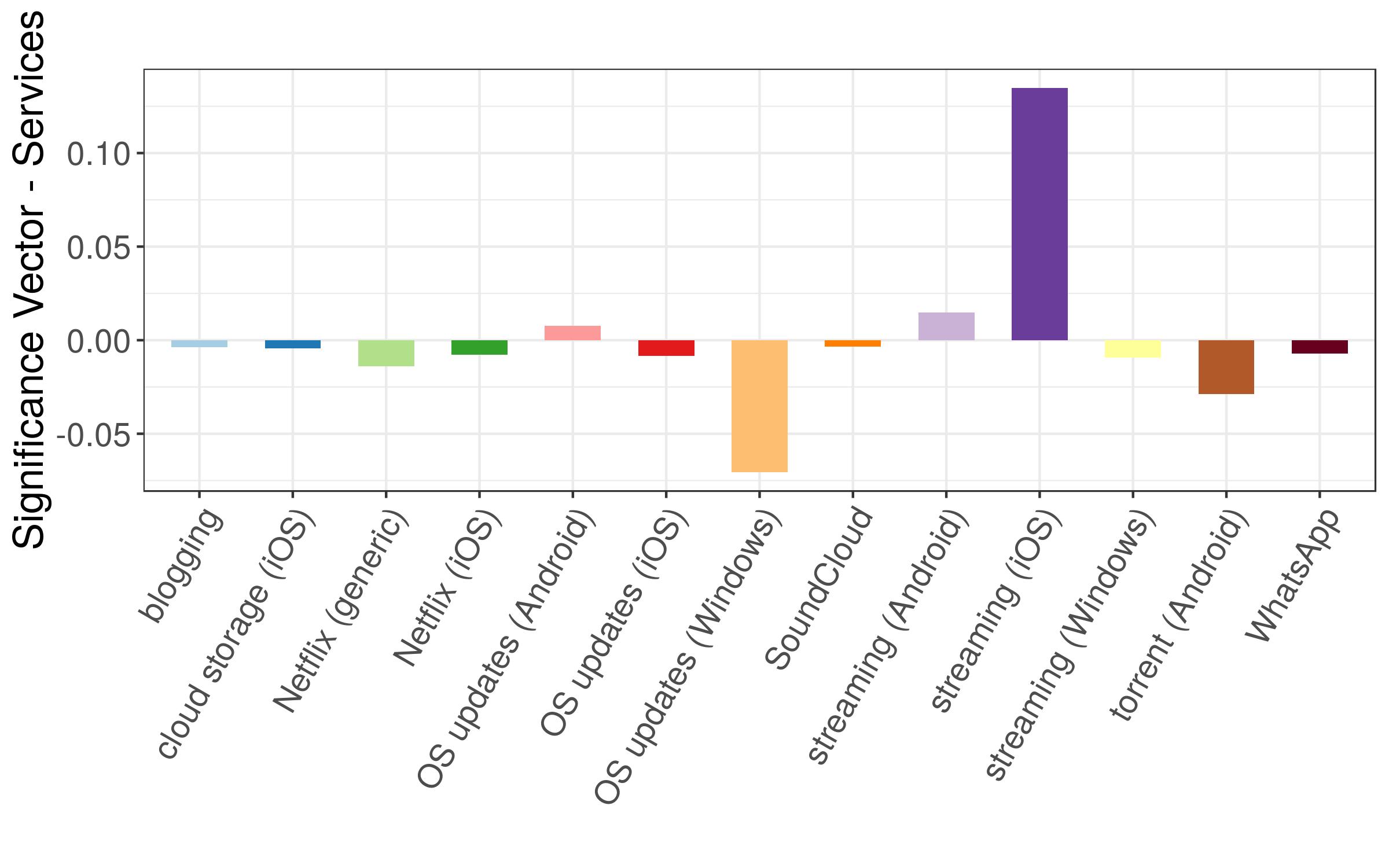}}\label{fig:1_PDF_clustered_difference_step_9_cl_3_logFALSE_13_communes_999}
	\hspace*{-2mm}
	\subfigure[Cluster 5]{\includegraphics[width=0.52\columnwidth,trim={0 25 0 0},clip]{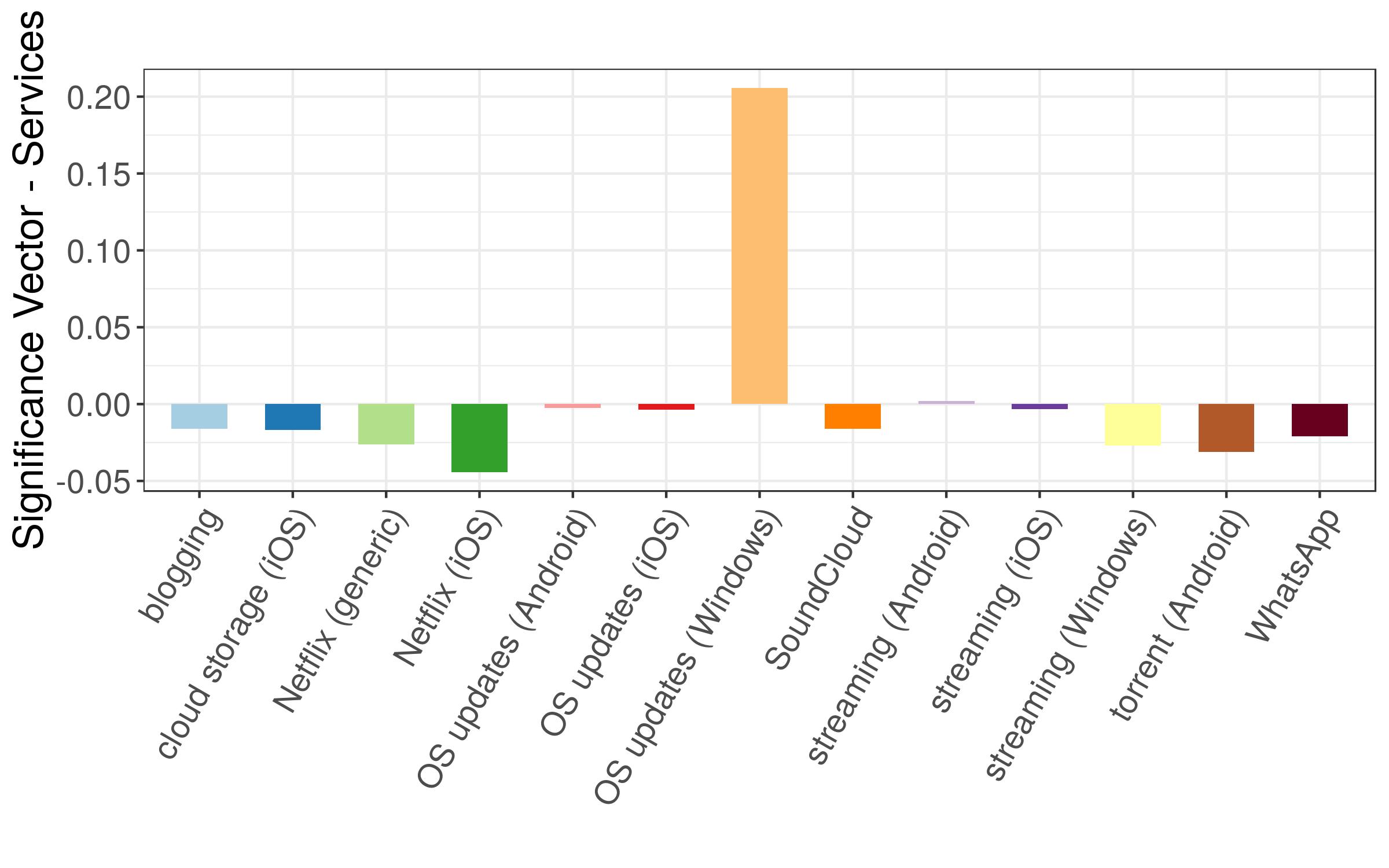}\label{fig:1_PDF_clustered_difference_step_9_cl_5_logFALSE_13_communes_999}}
	\hspace*{-2mm}
	\subfigure[Cluster 9]{\includegraphics[width=0.52\columnwidth,trim={0 25 0 0},clip]{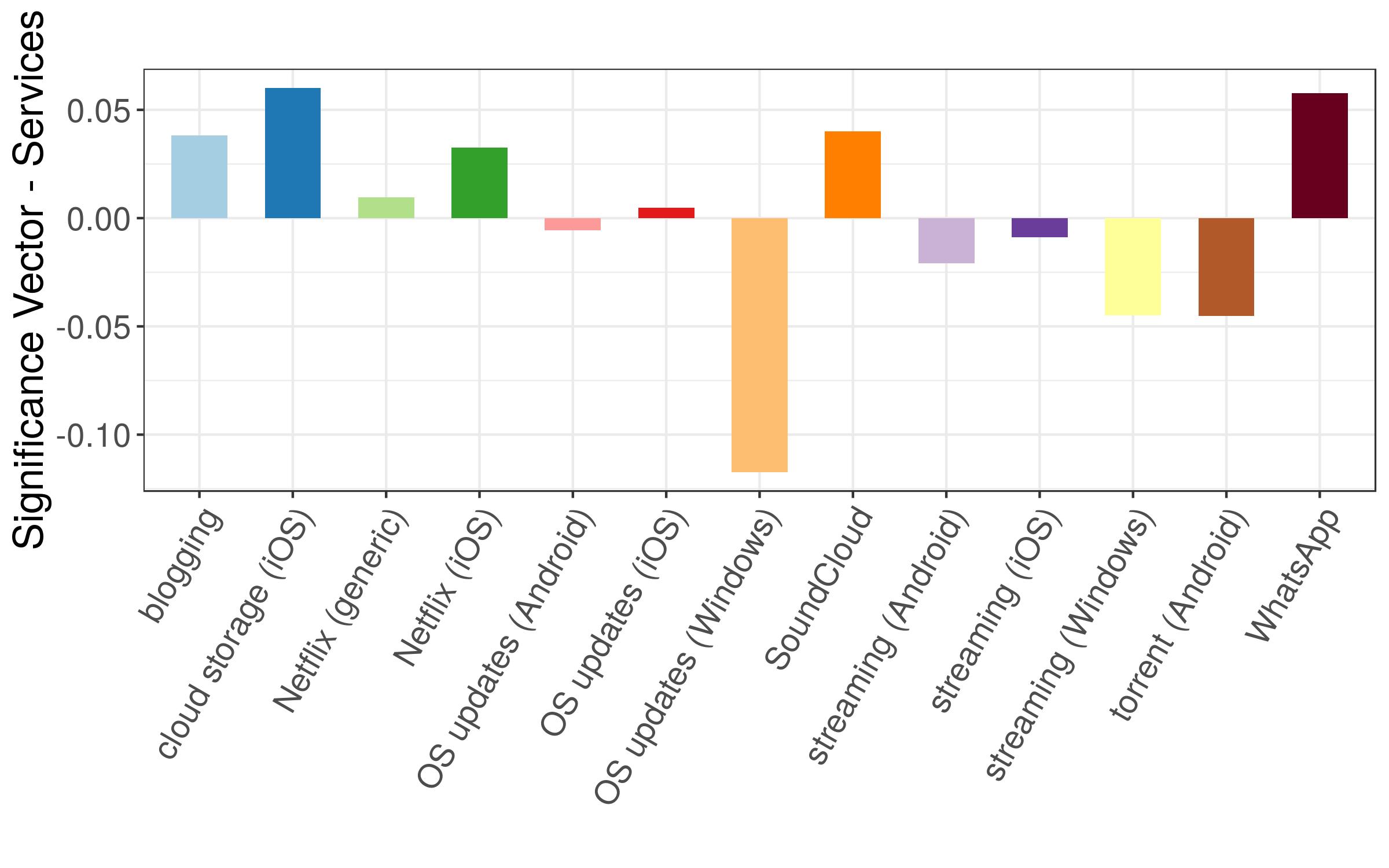}\label{fig:1_PDF_clustered_difference_step_9_cl_9_logFALSE_13_communes_999}}
	\vspace*{-12pt}
	\caption{Significance vectors $\rhov_k^{S'}(j)$ for the four clusters $k\in\mathcal{K}$ in Fig.\,\ref{fig:pdf}.}
	\label{fig:sig-s}
	\vspace*{-8pt}
\end{figure*}
\begin{figure*}[tb]
	\centering
	\subfigure[Cluster 2]{\includegraphics[width=0.48\columnwidth,trim={0 5 0 0},clip]{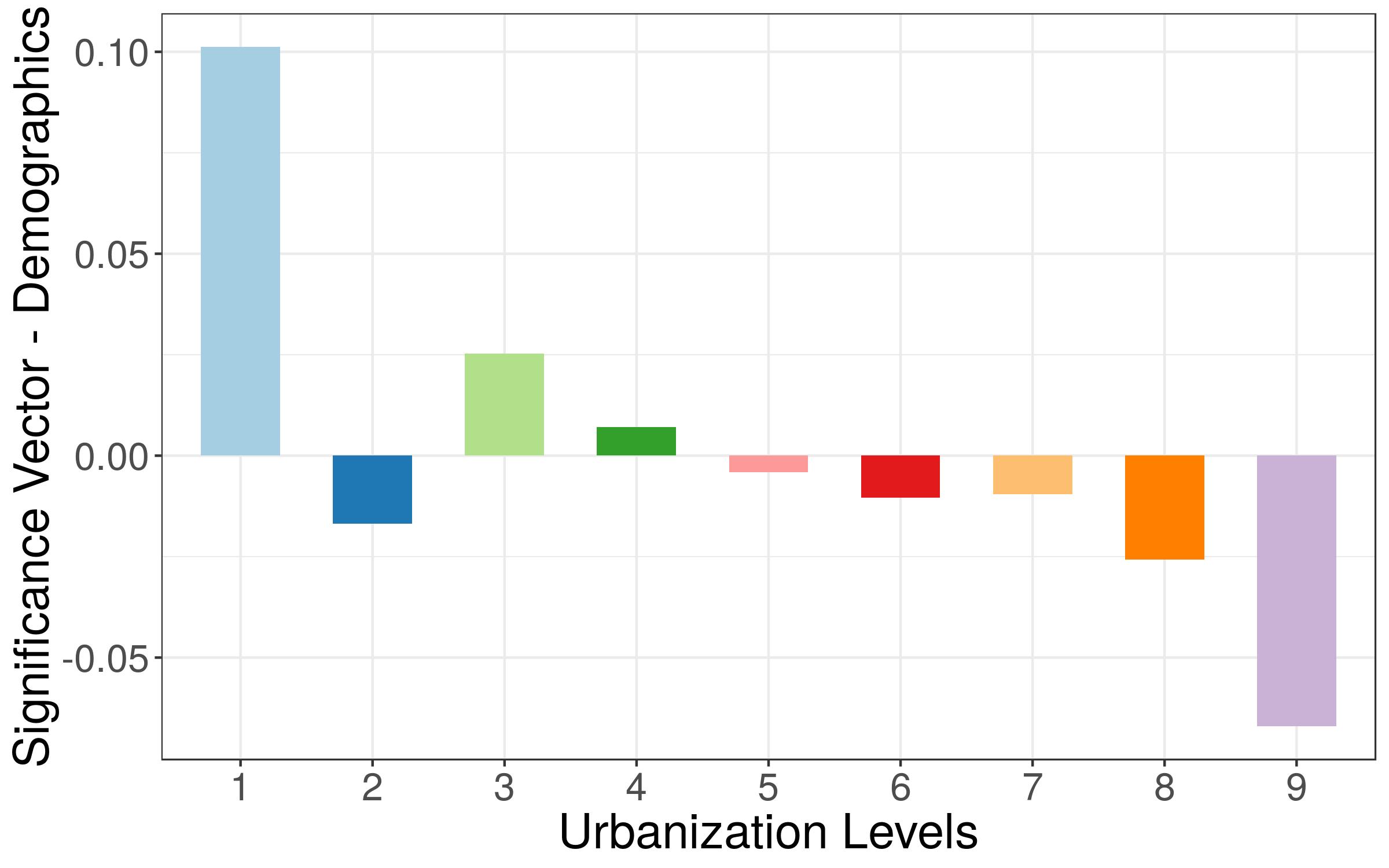}\label{fig:1_PDF_minus_avg_urbn_step_9_cl_2_13_communes_999}}
	\subfigure[Cluster 3]{\includegraphics[width=0.48\columnwidth,trim={0 5 0 0},clip]{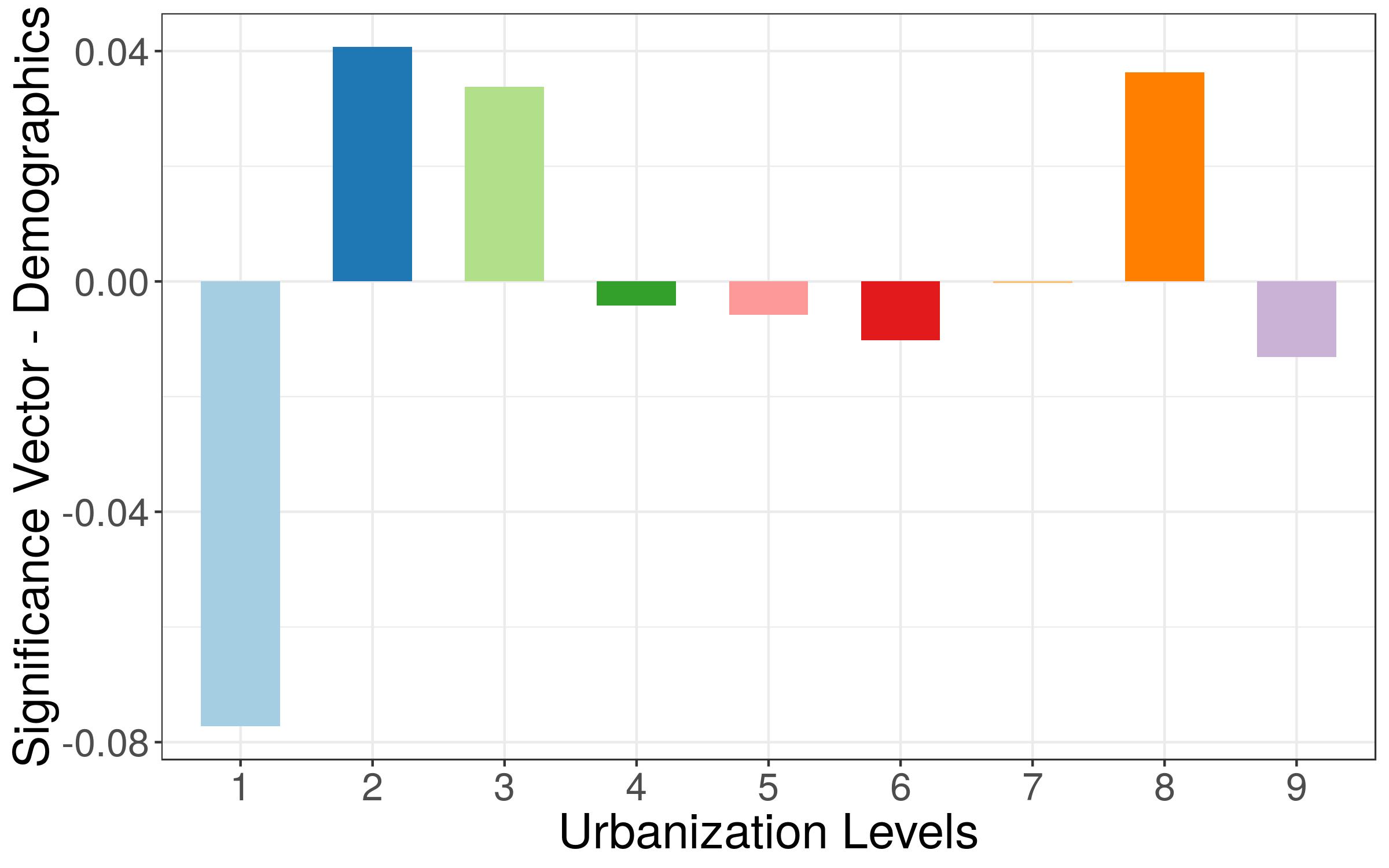}}\label{fig:1_PDF_minus_avg_urbn_step_9_cl_3_13_communes_999}
	\subfigure[Cluster 5]{\includegraphics[width=0.48\columnwidth,trim={0 5 0 0},clip]{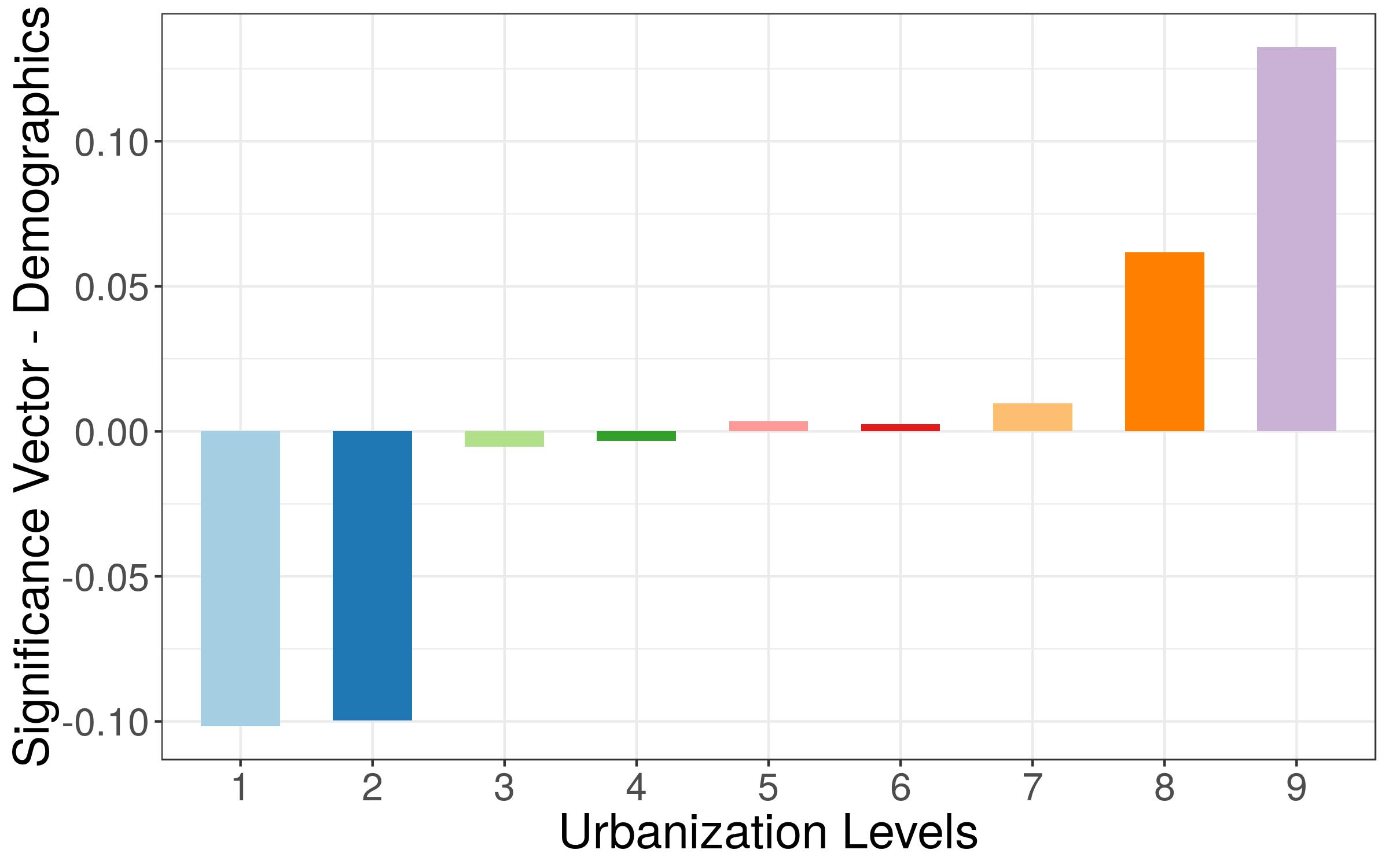}\label{fig:1_PDF_minus_avg_urbn_step_9_cl_5_13_communes_999}}
	\subfigure[Cluster 9]{\includegraphics[width=0.48\columnwidth,trim={0 5 0 0},clip]{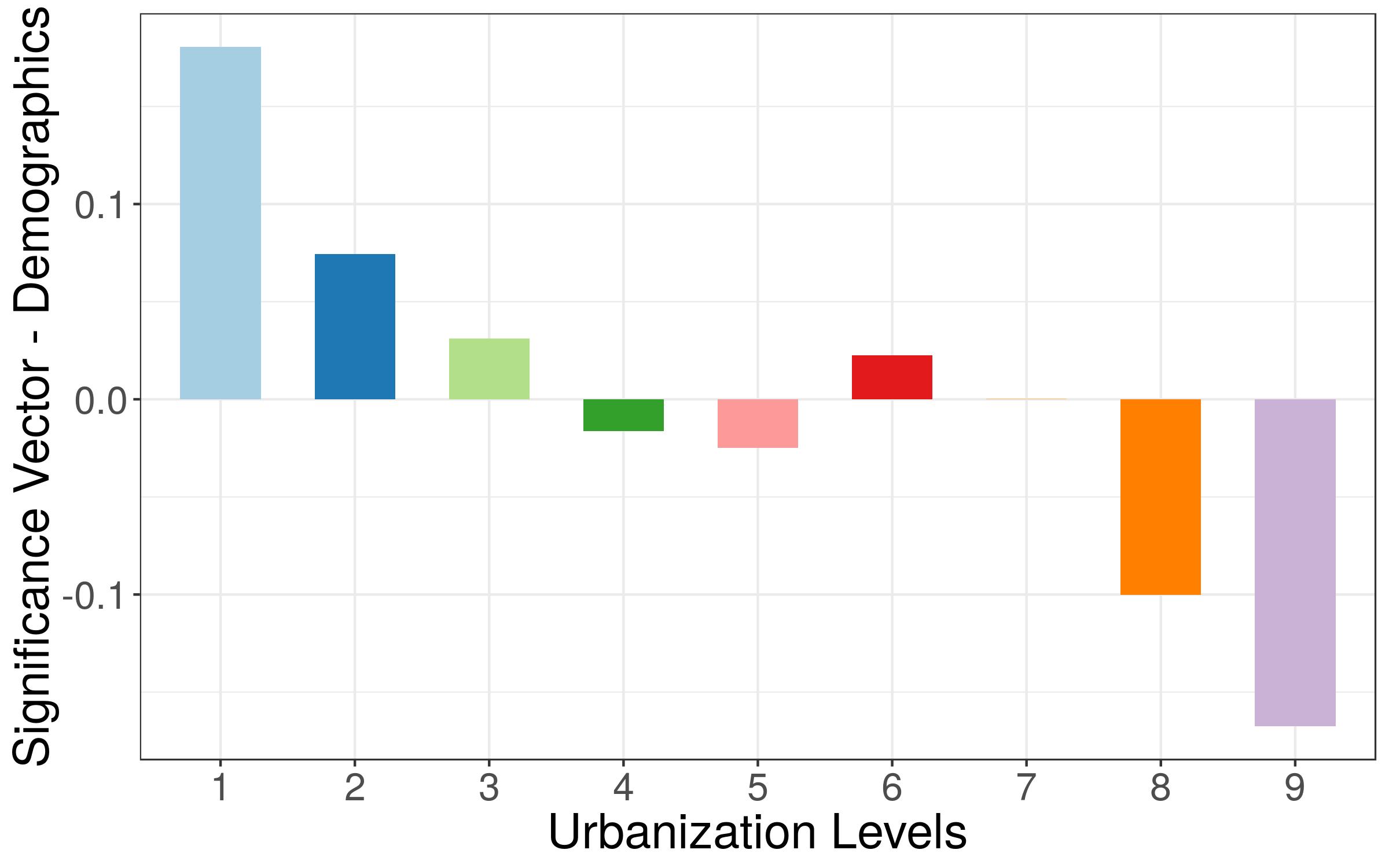}\label{fig:1_PDF_minus_avg_urbn_step_9_cl_9_13_communes_999}}
	\vspace*{-10pt}
	\caption{Significance vectors $\rhov_k^D(d)$
		for the four clusters $k\in\mathcal{K}$ in Fig.\,\ref{fig:pdf}. We consider urbanization levels as the levels $d\in\mathcal{D}$.
	}
	\label{fig:sig-d}
	\vspace*{-8pt}
\end{figure*}

We now define \emph{significance vectors} $\rhov_k^D(d)$ and $\rhov_k^{S'}(j)$ from the distributions $p_{D|K}(d|k)=\rho_{d,k}$ and $p_{S'|K}(j|k)=\rho_{j,k}$, as:
\[
\rhov_k^D(d) = \rho_{d,k}-\frac{1}{N_K-1}\sum_{k'\in\mathcal{K},k'\neq k}\rho_{d,k'}, \quad \forall k\in\mathcal{K},
\]

\[
\rhov_k^{S'}(j) = \rho_{j,k}-\frac{1}{N_K-1}\sum_{k'\in\mathcal{K},k'\neq k}\rho_{j,k'}, \quad \forall k\in\mathcal{K}.
\]

The significance vector $\rhov_k^D(d)$ (respectively, $\rhov_k^{S'}(j)$) associated to cluster $k$ thus assigns a weight bounded in $[-1,1]$ to each demographics level (respectively, service). Such weight indicates how much the fraction of communes in the clusters associated to one demographics level (respectively, the demand for a specific service in the cluster) differ, positively or negatively, from the average across all clusters.
Example illustrations are provided in Fig.\,\ref{fig:sig-s} and Fig.\,\ref{fig:sig-d}. Fig.\,\ref{fig:sig-s} shows the significance vectors $\rhov_k^{S'}(j)$ derived from the conditional service distributions at the four representative clusters in Fig.\,\ref{fig:pdf}.
Fig.\,\ref{fig:sig-d} portrays the significance vectors $\rhov_k^D(d)$ of the same four clusters, with respect to urbanization levels. We can observe, for instance, that clusters 2 and 9 show similar patterns in $\rhov_k^{S'}(j)$ and are much more present in dense urban areas. Instead, cluster 5 is characterized by very high incidence of Windows Mobile updates, as well as by a striking presence in rural regions.

\begin{figure*}[tb]
	\centering
	\hspace*{-5pt}
	\includegraphics[height=0.36\columnwidth,trim={0 0 0 0},clip]{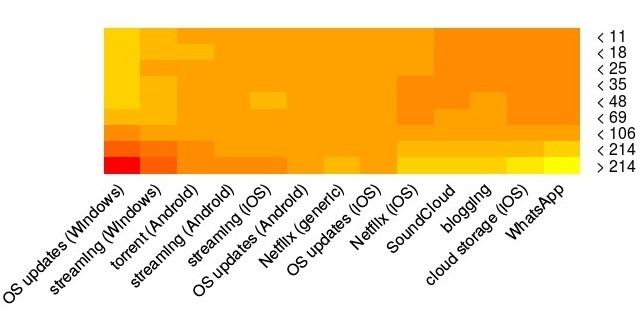}
	\hspace*{-6pt}
	\includegraphics[height=0.36\columnwidth,trim={0 0 0 0},clip]{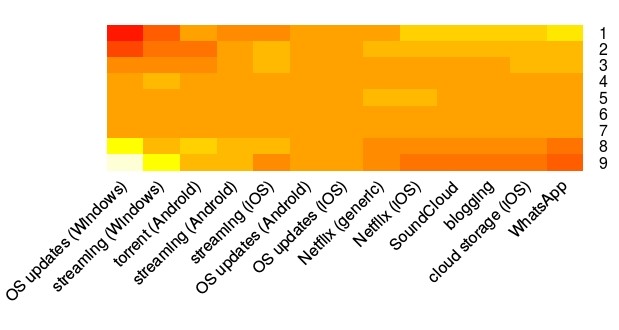}
	\hspace*{15pt}
	\includegraphics[height=0.33\columnwidth]{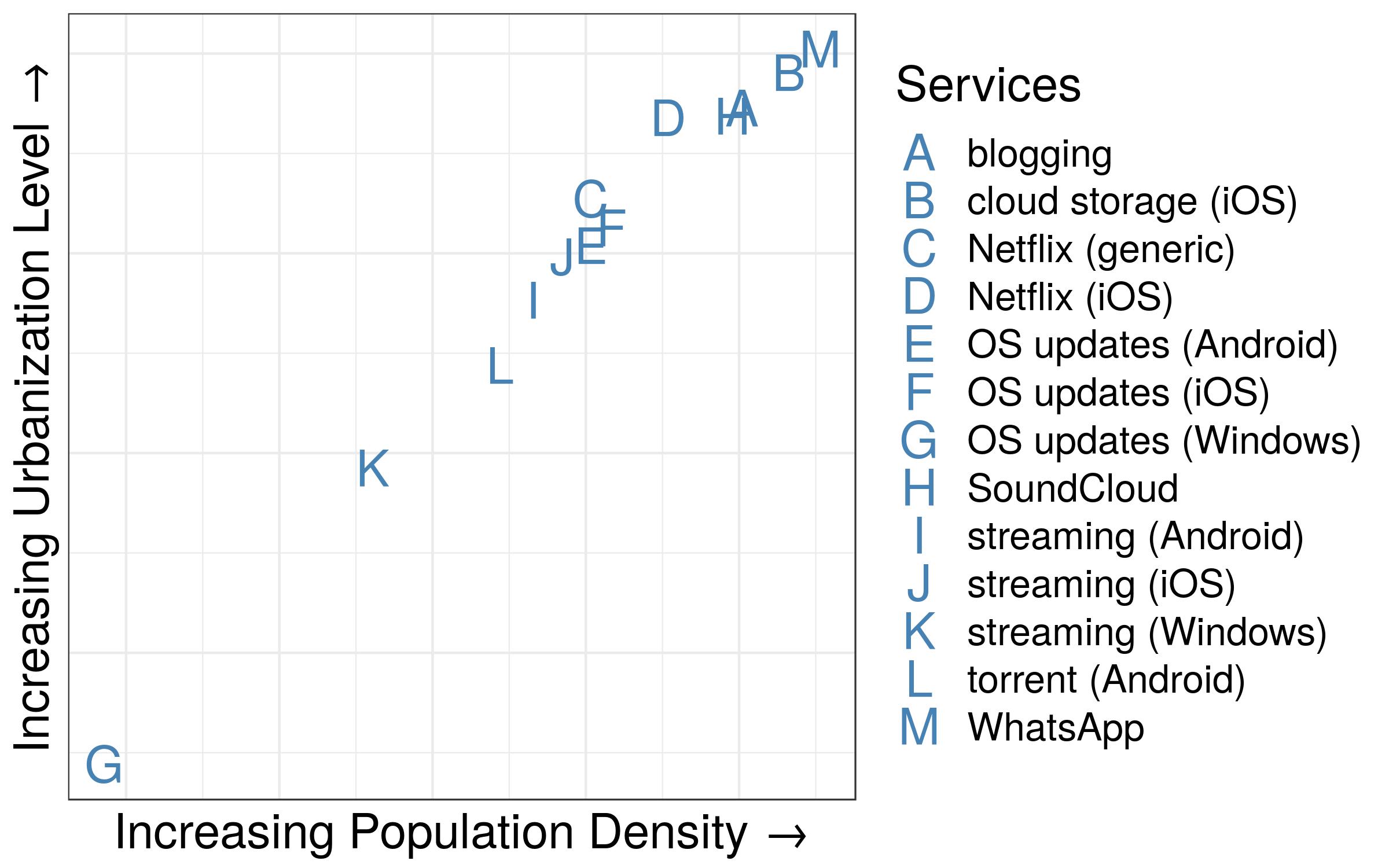}
	\vspace*{-8pt}
	\caption{Left, middle: incidence matrices $\Mm(d,j)$, as heatmaps where light (respectively, dark) colors denote high (respectively, low) values. Demographics levels in $\mathcal{D}$ are derived from discretized population density in terms of number of people per square kilometer (left) and urbanization levels (middle). Services are from the informative set $\mathcal{S}'$. Right: relative positioning of informative mobile services in the space of population density and urbanization levels.}
	\label{fig:matrices}
	\vspace*{-12pt}
\end{figure*}

The extent of inter-dependency among services and demographics levels can be computed in a rigorous way for each cluster $k\in\mathcal{K}$, by means of the matrix multiplication $\rhov_k^D(d) \left[\rhov_k^{S'}(j)\right]^T$, where $[\cdot]^T$ is the matrix transposition operation. This returns an \emph{incidence matrix} $\Mm_k(d,j) \in \mathbb{R}^{N_D \times N_{S'}}$, where element $(d,j)$ is a value in $[-1,1]$ that indicates the peculiarity of demands for service $j$ in demographics level $d$, as conveyed by cluster $k$. Highly positive (resp., negative) values point at abnormally high (resp., low) incidence of the service in the demographics level.
Finally, it is possible to derive a single incidence matrix for the overall clustering $\mathfrak{K}$, by simply averaging over all clusters $k\in\mathcal{K}$, \ie $\Mm(d,j) = \frac{1}{N_{K_2}} \sum_{k\in\mathcal{K}} \Mm_k(d,j)$.

\subsubsection{Results and discussion}

Examples of the matrices $\Mm(d,j)$ obtained with the clustering $\mathfrak{K}$ portrayed in the left plot of Fig.\,\ref{fig:map9}, where $N_{K_2}=9$, are in the left and middle plots of Fig.\,\ref{fig:matrices}, for demographics levels obtained with population density percentiles and urbanization levels, respectively.
The matrices confirm that some informative services are prone to an higher-than-average use in areas characterized by specific demographics levels; for instance, Windows system updates and streaming services have higher incidence in rural areas, and a lower impact on urbanized areas; conversely, WhatsApp is widely adopted in metropolitan areas, but shows lower-than-average usage in the countryside.
Interestingly, the two metrics used to derive demographics levels, in the left and middle plots of Fig.\,\ref{fig:matrices}, appear to yield consistent views across services. We provide a joint representation in the right plot of Fig.\,\ref{fig:matrices}, where application $j\in\mathcal{S}'$ is located in the bidimensional space of population density and urbanization levels, by assigning to it coordinates $\frac{1}{N_D}\sum_{d\in\mathcal{D}}d\cdot\Mm(d,j)$, one for each notion of $\mathcal{D}$.
The plot confirms our observation of a clear association of specific informative services to different demographics, providing an interesting and consistent ranking of applications versus urbanization.

Specifically, it is apparent that people living in rural regions of France have a preference to use Windows Mobile devices. Automatic updates for such OS are especially characterizing of the spatial diversity, due to a lower overall traffic per user in rural regions, which makes background traffic stand out. Instead, inhabitants of French metropolitan areas prefer Apple iPhones, as iOS-only services have a higher incidence than normal in cities. Residents in French cities also display a remarkable tendency to significantly use WhatsApp, a popular messaging application, and long-lived streaming services such as Netflix.

\begin{figure}[tb]
	\centering
	\includegraphics[height=0.34\columnwidth,trim={100 20 200 0},clip]{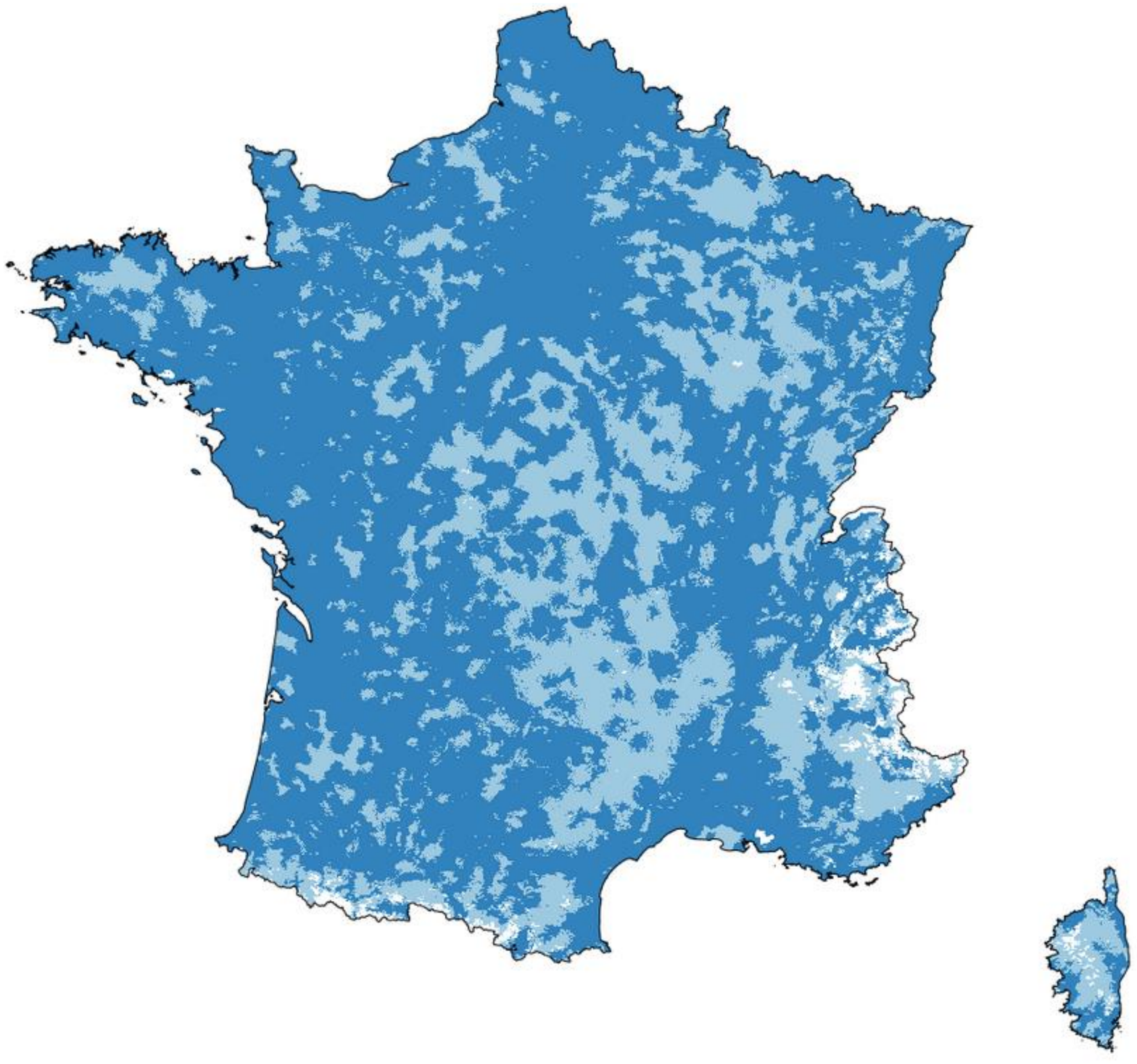}
	\hspace*{5pt}
	\raisebox{0pt}{
		\includegraphics[height=0.30\columnwidth,trim={0 0 5 0},clip]{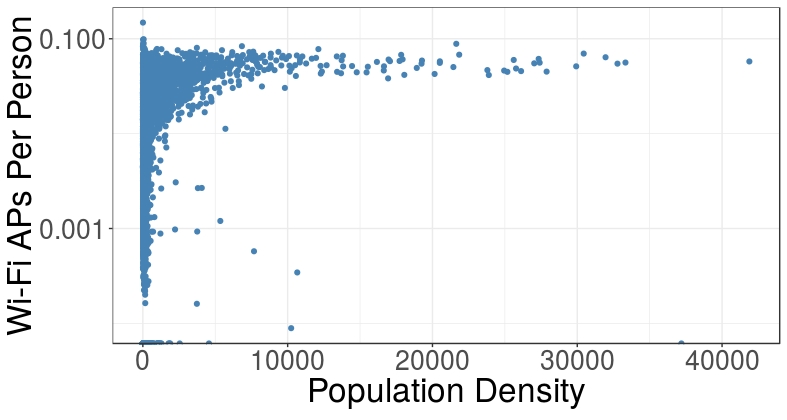}
	}
	\vspace*{-10pt}
	\caption{Left: 3G (light) and 4G (dark) coverage in France. Right: Scatterplot of the population density and Wi-Fi access point per person recorded in all communes of France.}
	\label{fig:access}
	\vspace*{-12pt}
\end{figure}

We underscore that the diversity of mobile service usage observed above is not an artifact of the different availability of radio access technolologies in urban and rural areas of France. The observed diversity of OS-specific traffic is a first evidence: there is no reason why lower or higher mobile datarates should affect adoption of Windows Mobile devices rather than Apple iPhones. To further prove our point, we leverage 2G, 3G, and 4G coverage maps provided by ARCEP, the French agency in charge of regulating telecommunications in France%
\footnote{\texttt{https://www.data.gouv.fr/fr/datasets/monreseaumobile/.}},
as well as data on the deployment of 1.8 million Wi-Fi home access points owned by Free, an incumbent Internet service provider in the country~\cite{rouveyrol15}. Both datasets refer to around the same period of the mobile service traffic collection.
The left plot in Fig.\,\ref{fig:access} shows that 4G coverage (dark blue) was already pervasive in France at the end of 2016: most of the national territory enjoyed broadband mobile access, including many regions tagged as suburban and rural by the demographics levels in Fig.\,\ref{fig:map9}. In the relatively small portion of geographic surface without 4G access, 3G was available, and 2G-only coverage areas were basically absent. Therefore, the cellular access technology cannot be considered as a discriminant for the geographical diversity of usages observed in our study for specific services.
Also, the right plot in Fig.\,\ref{fig:access} shows that Wi-Fi presence, measured in available access points per person, is largely uniform across communes; it ranges between 10 and 50 people per one Free Wi-Fi router, without any clear correlation with population density. The limited heterogeneity of Wi-Fi presence lets us conclude that also Wi-Fi access alone is insufficient to justify the differences in mobile service consumption across urban and rural areas that we find in the France scenario.

Overall, the results in this section lead to our final takeaway message: \textit{\textbf{there exist clear interplays between the usage of a specific set of mobile services and the demographics features of the territory, \ie people tend to consume differently some applications in cities and in the countryside}}. Such relationships are time-invariant: tests not detailed here due to space limitations return nearly identical results to those in Fig.\,\ref{fig:matrices} when disaggregating the data into night, morning, afternoon, and evening hours.

\section{Conclusions}

We unveil that most mobile services are typically consumed in very similar ways across a whole country like France. This suggests that heterogeneity in mobile service usage is observable at citywide scale due to land use~\cite{furno17}, and at worldwide scale due to cultural and language differences~\cite{peltonen18}, but it is much weaker at nationwide scale. In the latter case, we show that only a few specific services display spatial diversity, in a way that is strongly linked to urbanization.
While our results are for France, the methods used to derive them are general. Also, they are ductile, which leaves space for fine tuning and improvements: for instance, the two-phase algorithm can accommodate any clustering technique as the $\texttt{\small cluster}$ procedure, including more complex solutions than a greedy strategy.
Our insights and approach can be useful in mobile networking (for infrastructure planning), sociology (to understand relationships between digital activity and social segregation), or urban planning (to correlate mobile service usage and city structures).

\vspace*{-8pt}
\begin{acks}
The authors would like to thank Dr. Mathieu Cunche for providing the Wi-Fi deployment data.
R. Singh is supported in part by a PhD studentship under the EPSRC Centre for Doctoral Training in Pervasive Parallelism at the University of Edinburgh. R. Singh and M. Marina are also supported in part by The Alan Turing Institute through the PhD Enrichment scheme and Turing Fellowship, respectively.
The work of M. Fiore was partially supported by the European Union Horizon 2020 Framework Programme under REA grant agreement no.778305 DAWN4IoE.
\end{acks}

\bibliographystyle{ACM-Reference-Format}
\balance 
\bibliography{main}

\end{document}